\documentclass[12pt,twoside, a4paper]{article}
\pdfoutput=1
\def\pd{\partial}
\def\mc{\mathcal}

\usepackage[dvips]{graphicx}
\usepackage{amssymb}
\usepackage{amssymb,amsmath}
\usepackage{graphicx}
\usepackage{dsfont}
\usepackage{caption}
\usepackage{subcaption}
\usepackage{mathtools}
\usepackage{verbatim}
\usepackage{graphicx}
\usepackage{multirow}
\usepackage[outline]{contour}
\usepackage{xcolor,colortbl}
\usepackage{pdflscape}
\input{epsf.sty}  
\begin{document}
\begin{center}
\LARGE{\textbf{New supersymmetric Janus solutions from $N=4$ gauged supergravity}}
\end{center}
\vspace{1 cm}
\begin{center}
\large{\textbf{Tissana Assawasowan}$^a$ and \textbf{Parinya Karndumri}$^b$}
\end{center}
\begin{center}
String Theory and Supergravity Group, Department
of Physics, Faculty of Science, Chulalongkorn University, 254 Phayathai Road, Pathumwan, Bangkok 10330, Thailand
\end{center}
E-mail: $^a$tissana.a@hotmail.com\\
E-mail: $^b$parinya.ka@hotmail.com 
\vspace{1 cm}\\
\begin{abstract}
We study $N=4$ gauged supergravity with $SO(4)\times SO(4)$ gauge group in the presence of symplectic deformations and find new classes of Janus solutions preserving $N=1$ and $N=2$ supersymmetries. The $N=2$ solutions preserve $SO(2)\times SO(2)\times SO(2)\times SO(2)$ symmetry and interpolate between $N=4$ supersymmetric $AdS_4$ vacua with $SO(4)\times SO(4)$ symmetry. These correspond holographically to $N=(2,0)$ two-dimensional conformal defects within the dual $N=4$ CSM theories with $SO(4)\times SO(4)$ symmetry. The $N=1$ solutions contain two families of Janus configurations, one interpolating between $N=4$ $AdS_4$ vacua with $SO(4)\times SO(4)$ symmetry and the other interpolating between $N=4$ $AdS_4$ vacua with $SO(3)\times SO(3) \times SO(3)$ symmetry. These respectively describe $N=(1,0)$ conformal defects in $N=4$ CSM theories with $SO(4)\times SO(4)$ and $SO(3)\times SO(3) \times SO(3)$ symmetries. The latter give the first example of Janus solutions involving non-trivial $AdS_4$ vacua in addition to the trivial $SO(4)\times SO(4)$ critical point at the origin of the scalar manifold within the framework of $N=4$ gauged supergravity.
\end{abstract}
\newpage
\section{Introduction}
Janus configurations are solutions of gauged supergravity theories in the form of $AdS$-sliced (curved) domain walls interpolating between $AdS$ vacua. According to the AdS/CFT correspondence \cite{maldacena,Gubser_AdS_CFT,Witten_AdS_CFT}, these solutions holographically describe conformal interfaces or defects within the dual conformal field theories \cite{Bak_Janus}, see also \cite{Freedman_Janus,DHoker_Janus,Witten_Janus,Freedman_Holographic_dCFT}. These defects break the conformal symmetry of the bulk SCFT to that on the codimension-one defects by some position-dependent operators, see \cite{Gaunlett_spatial_mass1,Gaunlett_spatial_mass2} for recent results. For almost twenty years since the first Janus solution of \cite{Bak_Janus}, a large number of Janus solutions has been found in gauged supergravities in various space-time dimensions with different numbers of supersymmetries, see \cite{5D_Janus_CK,5D_Janus_DHoker1,
5D_Janus_DHoker2,5D_Janus_Suh,Bobev_5D_Janus1,Bobev_5D_Janus2,warner_Janus,N3_Janus,tri-sasakian-flow,orbifold_flow,Minwoo_4DN8_Janus,Kim_Janus,N5_flow,N6_flow,3D_Janus_de_Boer,3D_Janus_Bachas,3D_Janus_Bak,half_BPS_AdS3_S3_ICFT,exact_half_BPS_string,multi_face_Janus,
4D_Janus_from_11D,6D_Janus,3D_Janus,N8_omega_Janus,N4_Janus,ISO7_Janus,3D_Janus2} for an incomplete list. 
\\
\indent In this paper, we are interested in supersymmetric Janus solutions from symplectically deformed $N=4$ gauged supergravity with $SO(4)\times SO(4)$ gauge group. The $N=4$ gauged supergravity coupled to $n$ vector multiplets has been constructed in the embedding tensor formalism in \cite{N4_gauged_SUGRA}, see \cite{Eric_N4_4D,de_Roo_N4_4D,N4_Wagemans} for earlier construction, and possible symplectic deformations have been considered in \cite{Inverso_symplectic} extending the construction of $\omega$-deformed $SO(8)$ maximal gauged supergravity \cite{omega_N8_1,omega_N8_2,omega_Range1,omega_deWit} to lower numbers of supersymmetry. As pointed out in \cite{Inverso_symplectic}, for $N=4$ gauged supergravity with $SO(4)\times SO(4)\sim SO(3)\times SO(3)\times SO(3)\times SO(3)$ gauge group, there can be four deformation parameters or electric-magnetic phases for each $SO(3)$ factor. The first two $SO(3)$ factors are embedded in $SO(6)_R\sim SU(4)_R$ R-symmetry of $N=4$ supersymmetry. One of the phases for this $SO(3)\times SO(3)$ can be set to zero by $SL(2,\mathbb{R})$ transformations of the global symmetry $SL(2,\mathbb{R})\times SO(6,n)$ while the other gives equivalent gauged supergravities for any non-vanishing values and can be set to $\frac{\pi}{2}$. The phases of the remaining two $SO(3)$ factors embedded in the $SO(n)$ symmetry of the matter vector multiplets are independent deformation parameters in constrast to a single phase $\omega$ of the maximal $SO(8)$ gauged supergravity. The vacuum structure of the symplectically deformed $SO(4)\times SO(4)$ gauged supergravity has recently been investigated in \cite{N4_omega_flow} in which a large number of holographic RG flow solutions has also been given. In this paper, we will look for supersymmetric Janus solutions in this gauged supergravity. 
\\
\indent The study of Janus solutions in $N=4$ gauged supergravity has first appeared in \cite{tri-sasakian-flow} in which a number of singular Janus solutions, interpolating between singular geometries, has been given. The $N=4$ gauged supergravity in this case is obtained from a truncation of eleven-dimensional supergravity on a tri-sasakian manifold resulting in a non-semisimple $SO(3)\ltimes (\mathbf{T}^3,\widehat{\mathbf{T}}^3)$ gauge group. In addition, a regular Janus solution interpolating between the trivial $AdS_4$ vacua in $N=4$ gauged supergravity with $ISO(3)\times ISO(3)$ gauge group, obtained from a non-geometric compactification of type IIB theory, has been given in \cite{orbifold_flow}. In this case, the solution only involves scalar fields from the gravity multiplet. Both of these $N=4$ gauged supergravities admit only one supersymmetric $AdS_4$ vacuum at the origin of the scalar manifold. Therefore, Janus solutions involving more than one critical point are not possible. 
\\
\indent Regular Janus solutions, with non-vanishing scalars from both gravity and vector multiplets, in the framework of matter-coupled $N=4$ gauged supergravity with $SO(4)\times SO(4)$ gauge group have appeared only recently in \cite{N4_Janus}. This $N=4$ gauged supergravity admit a number of supersymmetric $AdS_4$ vacua \cite{4D_N4_flows} and can be obtained from the symplectically deformed $SO(4)\times SO(4)$ gauged supergravity mentioned above for a particular choice of electric-magnetic phases, two of the phases vanishing and the other two equal $\frac{\pi}{2}$. However, the solutions found in \cite{N4_Janus} are obtained only in $SO(2)\times SO(2)\times SO(3)\times SO(2)$ subtruncation of $SO(2)\times SO(2)\times SO(2)\times SO(2)$ scalar sector in which only the trivial $SO(4)\times SO(4)$ $AdS_4$ critical point appears. Accordingly, Janus solutions in \cite{N4_Janus} only interpolate between the trivial $SO(4)\times SO(4)$ critical points as well.
\\
\indent In the present paper, we will extend this study in two main aspects. We first look at $N=2$ Janus solutions in the full $SO(2)\times SO(2)\times SO(2)\times SO(2)$ scalar sector. Although no free deformation parameters appear in this sector as shown in \cite{N4_omega_flow}, we do find a number of new $N=2$ Janus solutions with $SO(2)\times SO(2)\times SO(2)\times SO(2)$ symmetry generalizing the results of \cite{N4_Janus}. Secondly, we will consider $SO(3)_{\textrm{diag}}\times SO(3)$ sector which, in addition to the trivial $SO(4)\times SO(4)$ critical point, admits two non-trivial $N=4$ $AdS_4$ critical points \cite{N4_omega_flow}. We will find $N=1$ supersymmetric Janus solutions that are dependent on the electric-magnetic phases. Moreover, we also find a new family of $N=1$ Janus solutions interpolating between non-trivial $AdS_4$ critical points. To the best of our knowledge, these are the first Janus solutions that involve non-trivial $AdS_4$ critical points in the framework of four-dimensional $N=4$ gauged supergravity. Although a large number of Janus solutions of this type can be found in the maximal gauged supergravity, see for example \cite{warner_Janus,Minwoo_4DN8_Janus,N8_omega_Janus,ISO7_Janus}, apart from the solutions in three-dimensional $N=8$ gauged supergravity studied recently in \cite{3D_Janus2}, no such solutions have been found within half-maximal gauged supergravities in higher dimensions to date. We hope the result of this paper would constitute the first step to fill this gap.  
\\
\indent The paper is organized as follows. In section \ref{N4_SUGRA},
we review the structure of four-dimensional $N=4$ gauged supergravity with symplectically deformed $SO(4)\times SO(4)$ gauge group. We then set up BPS equations within $SO(2)\times SO(2)\times SO(2)\times SO(2)$ and $SO(3)_{\textrm{diag}}\times SO(3)$ truncations and find a number of Janus solutions preserving $N=2$ and $N=1$ supersymmetries in sections \ref{N2Janus} and \ref{N1Janus}, respectively. We end the paper with some conclusions and comments in section \ref{conclusion}.
\section{Matter-coupled $N=4$ gauged supergravity}\label{N4_SUGRA} 
In this section, we give a brief review of $N=4$ gauged supergravity coupled to vector multiplets in the embedding tensor formalism constructed in \cite{N4_gauged_SUGRA}. The gravity and vector multiplets read
\begin{equation}
(e^{\hat{\mu}}_\mu,\psi^i_\mu,A_\mu^m,\chi^i,\tau)
\end{equation}
and
\begin{equation}
(A^a_\mu,\lambda^{ia},\phi^{ma}).
\end{equation} 
The bosonic component fields from the gravity and $n$ vector multiplets are given by the graviton $e^{\hat{\mu}}_\mu$, $6+n$ vector fields $A^{+M}=(A^m_\mu,A^a_\mu)$, a complex scalar $\tau$ containing the dilaton $\phi$ and the axion $\chi$ parametrizing $SL(2,\mathbb{R})/SO(2)$ coset, and $6n$ scalars $\phi^{ma}$ parametrizing $SO(6,n)/SO(6)\times SO(n)$ coset. Indices $\mu,\nu,\ldots =0,1,2,3$ and $\hat{\mu},\hat{\nu},\ldots=0,1,2,3$ denote respectively space-time and tangent space (flat) indices while $m,n=1,\ldots, 6$ and $i,j=1,2,3,4$ indices describe fundamental representations of $SO(6)_R$ and $SU(4)_R$ R-symmetry. The $n$ vector
multiplets are labeled by indices $a,b=1,\ldots, n$. The vector fields $A^{+M}$ and the magnetic dual $A^{-M}$ form a doublet under $SL(2,\mathbb{R})$ and will be collectively denoted by $A^{\alpha M}$, $\alpha=(+,-)$. 
\\
\indent The fermionic fields contain four gravitini $\psi^i_\mu$, four spin-$\frac{1}{2}$ fields $\chi^i$ and $4n$ gaugini $\lambda^{ia}$. These fields and supersymmetry parameters are subject to the chirality projections
\begin{equation}
\gamma_5\psi^i_\mu=\psi^i_\mu,\qquad \gamma_5\chi^i=-\chi^i,\qquad \gamma_5\lambda^{ia}=\lambda^{ia}
\end{equation}
and similarly for conjugate spinors
\begin{equation}
\gamma_5\psi_{\mu i}=-\psi_{\mu i},\qquad \gamma_5\chi_i=\chi_i,\qquad \gamma_5\lambda^a_i=-\lambda^a_i\, .
\end{equation}
\indent By using the complex scalar $\tau$ of the form
\begin{equation}
\tau=\chi+ie^\phi,
\end{equation}
we can write the coset representative for $SL(2,\mathbb{R})/SO(2)$ as
\begin{equation}
\mc{V}_\alpha=e^{\frac{\phi}{2}}\left(
                                         \begin{array}{c}
                                           \chi+ie^{\phi} \\
                                           1 \\
                                         \end{array}
                                       \right).
\end{equation}
Similarly, the $6n$ vector multiplet scalars $\phi^{ma}$ can be described by the coset representative 
\begin{equation}
\mc{V}_M^{\phantom{M}A}=(\mc{V}_M^{\phantom{M}m},\mc{V}_M^{\phantom{M}a}).
\end{equation}
We have decomposed the $SO(6)\times SO(n)$ index as $A=(m,a)$. We also note that the matrix $\mc{V}_M^{\phantom{M}A}$ satisfies the relation
\begin{equation}
\eta_{MN}=-\mc{V}_M^{\phantom{M}m}\mc{V}_N^{\phantom{M}m}+\mc{V}_M^{\phantom{M}a}\mc{V}_N^{\phantom{M}a}
\end{equation}
with $\eta_{MN}=\textrm{diag}(-1,-1,-1,-1,-1,-1,1,\ldots,1)$ being the $SO(6,n)$ invariant tensor. The inverse of $\mc{V}_M^{\phantom{M}A}$ will be denoted by ${\mc{V}_A}^M=({\mc{V}_m}^M,{\mc{V}_a}^M)$.
\\
\indent Gaugings of the matter-coupled $N=4$ supergravity are encoded in the components of the embedding tensor $\xi^{\alpha M}$ and $f_{\alpha MNP}$. We will consider only the gaugings with $\xi^{\alpha M}=0$ as required by the existence of supersymmetric $AdS_4$ vacua \cite{AdS4_N4_Jan}. In addition, we will also set all fermionic and vector fields to zero since supersymmetric Janus solutions involve only the metric and scalar fields. The bosonic Lagrangian can then be written as
\begin{equation}
e^{-1}\mc{L}=\frac{1}{2}R+\frac{1}{16}\pd_\mu M_{MN}\pd^\mu
M^{MN}-\frac{1}{4(\textrm{Im}\tau)^2}\pd_\mu \tau \pd^\mu \tau^*-V
\end{equation}
where $e=\sqrt{-g}$ is the vielbein determinant. The scalar potential is given by
\begin{eqnarray}
V&=&\frac{1}{16}\left[f_{\alpha MNP}f_{\beta
QRS}M^{\alpha\beta}\left[\frac{1}{3}M^{MQ}M^{NR}M^{PS}+\left(\frac{2}{3}\eta^{MQ}
-M^{MQ}\right)\eta^{NR}\eta^{PS}\right]\right.\nonumber \\
& &\left.-\frac{4}{9}f_{\alpha MNP}f_{\beta
QRS}\epsilon^{\alpha\beta}M^{MNPQRS}\right].
\end{eqnarray}
\indent The symmetric matrix $M_{MN}$, with the inverse $M^{MN}$, is defined by 
\begin{equation}
M_{MN}=\mc{V}_M^{\phantom{M}m}\mc{V}_N^{\phantom{M}m}+\mc{V}_M^{\phantom{M}a}\mc{V}_N^{\phantom{M}a}\, .
\end{equation}
The tensor $M^{MNPQRS}$ is obtained from
\begin{equation}
M_{MNPQRS}=\epsilon_{mnpqrs}\mc{V}_{M}^{\phantom{M}m}\mc{V}_{N}^{\phantom{M}n}
\mc{V}_{P}^{\phantom{M}p}\mc{V}_{Q}^{\phantom{M}q}\mc{V}_{R}^{\phantom{M}r}\mc{V}_{S}^{\phantom{M}s}\label{M_6}
\end{equation}
by raising indices with $\eta^{MN}$. The matrix $M^{\alpha\beta}$ is the inverse of the symmetric $2\times 2$ matrix $M_{\alpha\beta}$ defined by
\begin{equation}
M_{\alpha\beta}=\textrm{Re}(\mc{V}_\alpha\mc{V}^*_\beta).
\end{equation}
\indent We also need fermionic supersymmetry transformations 
\begin{eqnarray}
\delta\psi^i_\mu &=&2D_\mu \epsilon^i-\frac{2}{3}A^{ij}_1\gamma_\mu
\epsilon_j,\\
\delta \chi^i &=&-\epsilon^{\alpha\beta}\mc{V}_\alpha D_\mu
\mc{V}_\beta\gamma^\mu \epsilon^i-\frac{4}{3}iA_2^{ij}\epsilon_j,\\
\delta \lambda^i_a&=&2i\mc{V}_a^{\phantom{a}M}D_\mu
\mc{V}_M^{\phantom{M}ij}\gamma^\mu\epsilon_j-2iA_{2aj}^{\phantom{2aj}i}\epsilon^j
\end{eqnarray}
with the fermion shift matrices defined by
\begin{eqnarray}
A_1^{ij}&=&\epsilon^{\alpha\beta}(\mc{V}_\alpha)^*\mc{V}_{kl}^{\phantom{kl}M}\mc{V}_N^{\phantom{N}ik}
\mc{V}_P^{\phantom{P}jl}f_{\beta M}^{\phantom{\beta M}NP},\nonumber
\\
A_2^{ij}&=&\epsilon^{\alpha\beta}\mc{V}_\alpha\mc{V}_{kl}^{\phantom{kl}M}\mc{V}_N^{\phantom{N}ik}
\mc{V}_P^{\phantom{P}jl}f_{\beta M}^{\phantom{\beta M}NP},\nonumber
\\
A_{2ai}^{\phantom{2ai}j}&=&\epsilon^{\alpha\beta}\mc{V}_\alpha
{\mc{V}_a}^M{\mc{V}_{ik}}^N\mc{V}_P^{\phantom{P}jk}f_{\beta
MN}^{\phantom{\beta MN}P}\, .
\end{eqnarray}
The coset representative of the form $\mc{V}_M^{\phantom{M}ij}$ and ${\mc{V}_{ij}}^M$ are defined in terms of the 't Hooft
symbols $G^{ij}_m$ as
\begin{equation}
\mc{V}_M^{\phantom{M}ij}=\frac{1}{2}\mc{V}_M^{\phantom{M}m}G^{ij}_m
\end{equation}
and
\begin{equation}
{\mc{V}_{ij}}^M=-\frac{1}{2}{\mc{V}_{m}}^M(G^{ij}_m)^*\,
.
\end{equation}
The explicit representation of $G^{ij}_m$ used in this paper is the same as in \cite{N4_omega_flow}. Upper and lower $i,j,\ldots$ indices are related by complex conjugation as usual.

In this paper, we only consider $N=4$ gauged supergravity coupled to $n=6$ vector multiplets with $SO(4)\times SO(4)$ gauge group. By decomposing the $SO(6,6)$ fundamental index as $M=(\hat{m},\tilde{m},\hat{a},\tilde{a})$, for $\hat{m},\tilde{m},\hat{a},\tilde{a}=1,2,3$, we can write the embedding tensor for symplectically deformed $SO(4)\times SO(4)$ gauge group as 
\begin{eqnarray}
& &f_{+\hat{m}\hat{n}\hat{p}}=-g_0\cos\alpha_0 \epsilon_{\hat{m}\hat{n}\hat{p}}, \qquad f_{-\hat{m}\hat{n}\hat{p}}=g_0\sin\alpha_0 \epsilon_{\hat{m}\hat{n}\hat{p}}, \nonumber \\
& & f_{+\tilde{m}\tilde{n}\tilde{p}}=-g\cos\alpha \epsilon_{\tilde{m}\tilde{n}\tilde{p}},\qquad f_{-\tilde{m}\tilde{n}\tilde{p}}=g\sin\alpha \epsilon_{\tilde{m}\tilde{n}\tilde{p}},\nonumber \\ 
& &f_{+\hat{a}\hat{b}\hat{c}}=h_1\cos\beta_1 \epsilon_{\hat{a}\hat{b}\hat{c}}, \qquad f_{-\hat{a}\hat{b}\hat{c}}=h_1\sin\beta_1 \epsilon_{\hat{a}\hat{b}\hat{c}}, \nonumber \\
& & f_{+\tilde{a}\tilde{b}\tilde{c}}=h_2\cos\beta_2 \epsilon_{\tilde{a}\tilde{b}\tilde{c}},\qquad f_{-\tilde{a}\tilde{b}\tilde{c}}=h_2\sin\beta_2 \epsilon_{\tilde{a}\tilde{b}\tilde{c}}\, .
\end{eqnarray}      
These components of the embedding tensor have been given in \cite{Dibitetto_SL2_angle}, and we have rewritten them in the notation of \cite{Inverso_symplectic}. $f_{\pm \hat{m}\hat{n}\hat{p}}$ and $f_{\pm \tilde{m}\tilde{n}\tilde{p}}$ describe the embedding of the first $SO(4)\sim SO(3)\times SO(3)$ factor in $SO(6)_R$ R-symmetry. As previously mentioned, the constants $\alpha_0$ and $\alpha$ can be set to zero and $\frac{\pi}{2}$, respectively. $g_0$, $g$, $h_1$ and $h_2$ are gauge coupling constants for the four $SO(3)$ factors. In subsequent sections, we will look for supersymmetric Janus solutions with different numbers of unbroken supersymmetries and residual symmetries. 

\section{$N=2$ supersymmetric Janus solutions}\label{N2Janus}
We begin with a truncation to scalars that are singlets of $SO(2)\times SO(2)\times SO(2)\times SO(2)$ subgroup of the $SO(4)\times SO(4)$ gauge group. We first choose an explicit form of $SO(6,6)$ generators in the fundamental representation as
\begin{equation}
(t_{MN})_P^{\phantom{P}Q}=2\delta^Q_{[M}\eta_{N]P}\, .
\end{equation}
The $SO(6,6)$ non-compact generators are accordingly given by
\begin{equation}
Y_{ma}=t_{m,a+6}\, .
\end{equation}
Follow \cite{N4_Janus}, the coset representative for $SO(2)\times SO(2)\times SO(2)\times SO(2)$ singlet scalars can be written as
\begin{equation}
\mc{V}=e^{\phi_1 Y_{33}}e^{\phi_2 Y_{36}}e^{\phi_3 Y_{63}}e^{\phi_4 Y_{66}}\, .\label{SO2_4_coset}
\end{equation}
The metric ansatz takes the form of the usual $AdS_3$-sliced domain walls
\begin{equation}
ds^2=e^{2A(r)}\left(e^{\frac{2\rho}{\ell}}dx^2_{1,1}+d\rho^2\right)+dr^2
\end{equation}
in which $\ell$ denotes the radius of the $AdS_3$ slices. $dx^2_{1,1}=\eta_{\alpha\beta}dx^\alpha dx^\beta$, $\alpha,\beta=0,1$, is the flat metric on two-dimensional Minkowski space.

All scalars $\phi_i$, $i=1,2,3,4$, together with the dilaton $\phi$ and the axion $\chi$ are allowed to depend only on $r$. The analysis of relevant BPS equations has already appeared in many places, see for example \cite{warner_Janus,N3_Janus}, so we will simply summarize the results. The supersymmetry transformations $\delta\psi^{i}_{\hat{\alpha}}$ give the following equation
\begin{equation}
A'^2=W^2-\frac{1}{\ell^2}e^{-2A}\label{dPsi_BPS_eq}
\end{equation}
while $\delta \psi^i_{\hat{\rho}}$ gives the Killing spinor of the form
\begin{equation}
\epsilon^{\hat{i}}=e^{\frac{\rho}{2\ell}}\tilde{\epsilon}^{\hat{i}} 
\end{equation}
for $\rho$-independent spinors $\tilde{\epsilon}^{\hat{i}}$. In equation \eqref{dPsi_BPS_eq}, $W=|\mc{W}|$, and the superpotential 
\begin{equation}
\mc{W}=\frac{2}{3}\hat{\alpha} 
\end{equation}
is obtained from the eigenvalue $\hat{\alpha}$ of $A^{ij}_1$ with the corresponding eigenvectors $\epsilon^{\hat{i}}$ identified with the Killing spinors. We use an index $\hat{i}$ to count the number of unbroken supersymmetry. 

With the projectors
\begin{equation}
\gamma_{\hat{r}}\epsilon^{\hat{i}}=e^{i\Lambda}\epsilon_{\hat{i}}\label{gamma_r_pro}
\end{equation}
and
\begin{equation}
\gamma_{\hat{\rho}}\epsilon^{\hat{i}}=i\kappa e^{i\Lambda}\epsilon_{\hat{i}}\label{gamma_rho_pro}
\end{equation}
with $\kappa^2=1$ and an $r$-dependent phase $\Lambda$, the Killing spinors can be determined from $\delta \psi^i_{\hat{r}}$ to be
\begin{equation}
\epsilon^{\hat{i}}=e^{\frac{A}{2}+\frac{\rho}{2\ell}+i\frac{\Lambda}{2}}\varepsilon^{(0){\hat{i}}}\, .
\end{equation}
The spinors $\varepsilon^{(0)\hat{i}}$ can (possibly) have an $r$-dependent phase and satisfy the following projection conditions
\begin{equation}
\gamma_{\hat{r}}\varepsilon^{(0){\hat{i}}}=\varepsilon^{(0)}_{\hat{i}}\qquad
\textrm{and}\qquad
\gamma_{\hat{\rho}}\varepsilon^{(0){\hat{i}}}=i\kappa\varepsilon^{(0)}_{\hat{i}}\,
.
\end{equation}
With all these results, the conditions $\delta \psi^i_{\hat{\alpha}}$ determine the explicit form of the phase $e^{i\Lambda}$ to be
\begin{equation}
e^{i\Lambda}=\frac{\mc{W}}{A'+\frac{i\kappa}{\ell}e^{-A}}=\frac{\mc{W}}{W^2}\left(A'-\frac{i\kappa}{\ell}e^{-A}\right)\,
.\label{complex_W_phase}
\end{equation}
With the projector \eqref{gamma_r_pro}, the variations $\delta\chi^{i}$ and $\delta\lambda^i_a$ lead to the BPS equations for scalars. Finally, we note that the sign factor $\kappa=\pm 1$ corresponds to chiralities of the Killing spinors on the two-dimensional defects.

For the $SO(2)\times SO(2)\times SO(2)\times SO(2)$ truncation, the $A_1^{ij}$ tensor takes the form, see more detail in \cite{N4_omega_flow},
\begin{equation}
A_1^{ij}=\textrm{diag}(\mc{A}_-,\mc{A}_+,\mc{A}_+,\mc{A}_-).
\end{equation}
Both of the eigenvalues lead to $N=2$ unbroken supersymmetry with the superpotential $\mc{W}_\mp=\frac{2}{3}\mc{A}_\mp$ and Killing spinors $\epsilon^{1,4}$ and $\epsilon^{2,3}$, respectively. Following \cite{N4_omega_flow}, we will set $\epsilon^2=\epsilon^3=0$ and choose the superpotential to be
\begin{eqnarray}
\mc{W}&=&\mc{W}_-\nonumber \\
&=&\frac{1}{2}e^{-\frac{\phi}{2}}\left[\cosh\phi_4[g\cosh\phi_3(e^\phi\sin\alpha+i\cos\alpha)-g_0\sinh\phi_1\sinh\phi_3] \right.\nonumber \\
& &\left. -g_0\cosh\phi_1(\cosh\phi_2+i\sinh\phi_2\sinh\phi_4)+ig\sin\alpha\cosh\phi_3\cosh\phi_4\chi
\right].\label{W_N2}
\end{eqnarray}
The scalar potential can be written in terms of the superpotential as
\begin{eqnarray}
V&=&-2G^{rs}\frac{\pd W}{\pd \Phi^r}\frac{\pd W}{\pd \Phi^s}-3W^2\nonumber \\
&=&-\frac{1}{4}e^{-\phi}[g^2(1+\cos2\alpha)+2g_0^2+2g^2\sin\alpha\chi(2\cos\alpha+\sin\alpha\chi)]-\frac{1}{2}e^\phi g^2\sin^2\alpha \nonumber \\
& &+2gg_0\sin\alpha \cosh\phi_1\cosh\phi_2\cosh\phi_3\cosh\phi_4\, .\label{Potential_SO2_4}
\end{eqnarray}
in which we have defined the scalars $\Phi^r=(\phi,\chi,\phi_1,\phi_2,\phi_3,\phi_4)$. $G^{rs}$ is the inverse of the scalar metric appearing in the scalar kinetic terms.

With the coset representative \eqref{SO2_4_coset}, the kinetic term for scalar fields is given by
\begin{eqnarray}
\mc{L}_{\textrm{kin}}&=&\frac{1}{2}G_{rs}{\Phi^r}'{\Phi^s}'\nonumber \\
&=&-\frac{1}{4}(\phi'^2+e^{-2\phi}\chi'^2)-\frac{1}{16}\left[6+\cosh2(\phi_2-\phi_3)\right. \nonumber \\
& &
\left.+\cosh2(\phi_2+\phi_3)+2\cosh2\phi_4(\cosh2\phi_2\cosh2\phi_3-1)\right]\phi'^2_1\nonumber \\
& &-\cosh\phi_2\cosh\phi_4\sinh\phi_3\sinh\phi_4\phi'_1\phi'_2-\cosh\phi_3\cosh\phi_4\sinh\phi_2\sinh\phi_4\phi'_1\phi'_3\nonumber \\
& &+\sinh\phi_2\sinh\phi_3\phi'_1\phi'_4-\frac{1}{2}\cosh^2\phi_4\phi'^2_2-\frac{1}{2}\cosh^2\phi_4\phi'^2_3-\frac{1}{2}\phi'^2_4\label{scalar_kin}
 \end{eqnarray}
from which we can determine the scalar metric $G_{rs}$ and its inverse $G^{rs}$. Since $G^{rs}$ will appear in the final form of the BPS equations, for later convenience, we will give its explicit form here
\begin{equation}
G^{rs}=\begin{pmatrix}	-2 & 0 & \mathbf{0}_{1\times 4} \\
						0 &-2e^{2\phi}  & \mathbf{0}_{1\times 4}\\
						    \mathbf{0}_{4\times 1} & \mathbf{0}_{4\times 1} & \widehat{G}^{\hat{r}\hat{s}}
			\end{pmatrix}
\end{equation}
with the $4\times 4$ symmetric matrix $\widehat{G}^{\hat{r}\hat{s}}$, for $\hat{r},\hat{s}=1,2,3,4$, given by
\begin{equation}
\widehat{G}^{\hat{r}\hat{s}}=\begin{pmatrix}	\square_1 & \Delta_1 & \Delta_2 & \Delta_3  \\
						\Delta_1 &\square_2  & \Delta_4 & \Delta_5 \\
						  \Delta_2&\Delta_4  & \square_3 & \Delta_6\\
						    \Delta_3 & \Delta_5 & \Delta_6 &\square_4 
			\end{pmatrix}
\end{equation}
and
\begin{eqnarray}
\square_1&=&-\textrm{sech}^2\phi_2\textrm{sech}^2\phi_3,\qquad \square_2=-\textrm{sech}^2\phi_3\textrm{sech}^2\phi_4-\tanh^2\phi_3,\nonumber \\
\square_3&=&\textrm{sech}^2\phi_2\tanh^2\phi_4-1,\qquad \square_4=-\frac{1}{2}\textrm{sech}^2\phi_2\textrm{sech}^2\phi_3(1+\cosh2\phi_2\cosh2\phi_3),\nonumber \\
\Delta_1&=& \textrm{sech}\phi_2\textrm{sech}\phi_3\tanh\phi_3\tanh\phi_4,\qquad
\Delta_2=\textrm{sech}\phi_2\textrm{sech}\phi_3\tanh\phi_2\tanh\phi_4,\nonumber \\
\Delta_3&=&-\textrm{sech}\phi_2\textrm{sech}\phi_3\tanh\phi_2\tanh\phi_3, \qquad 
\Delta_4=-\tanh\phi_2\tanh\phi_3\tanh^2\phi_4,\nonumber \\
\Delta_5&=&\tanh\phi_2\tanh^2\phi_3\tanh\phi_4,\qquad
\Delta_6=\tanh^2\phi_2\tanh\phi_3\tanh\phi_4\, .
\end{eqnarray}
\indent The scalar potential and superpotential admit one $AdS_4$ critical point at $\phi_1=\phi_2=\phi_3=\phi_4=0$ and  
\begin{equation}
\phi=\ln\left[-\frac{g_0}{g\sin\alpha}\right]\qquad \textrm{and} \qquad \chi=-\frac{\cos\alpha}{\sin\alpha}\, .
\end{equation} 
By shifting the dilaton and axion, or equivalently choosing $g_0=-g$ for $\alpha=\frac{\pi}{2}$, we can bring this critical point to the origin of the scalar manifold $SL(2,\mathbb{R})/SO(2)\, \times\, SO(6,6)/SO(6)\times SO(6)$ at which all scalars vanish. With this choice, the cosmological constant and $AdS_4$ radius are given by
\begin{equation} 
V_0=-3g^2\qquad\textrm{and}\qquad L=\sqrt{-\frac{3}{V_0}}=\frac{1}{g}
\end{equation}
in which we have taken $g>0$ without loss of generality. This critical point is invariant under the full $SO(4)\times SO(4)$ gauge symmetry and preserves $N=4$ supersymmetry.

Using the projector \eqref{gamma_r_pro} and the superpotential \eqref{W_N2}, we find that all the BPS conditions with $\epsilon^{2,3}=0$ lead to the following BPS equations 
\begin{eqnarray}
& &A'^2+\frac{1}{\ell^2} e^{-2A}=W^2,\\
& &\phi'=-4\frac{A'}{W}\frac{\pd W}{\pd \phi}-4e^\phi\frac{\kappa e^{-A}}{\ell W}\frac{\pd W}{\pd \chi},\\
& &\chi'=-4e^{2\phi}\frac{A'}{W}\frac{\pd W}{\pd \chi}+4e^\phi\frac{\kappa e^{-A}}{\ell W}\frac{\pd W}{\pd \phi},\\
& &\phi'_1=\widehat{G}^{1\hat{r}}\frac{A'}{W}\frac{\pd W}{\pd \widehat{\Phi}^{\hat{r}}}-2\textrm{sech}\phi_2\textrm{sech}\phi_3\textrm{sech}\phi_4\frac{\kappa e^{-A}}{\ell W}\frac{\pd W}{\pd \phi_3},\\
& &\phi'_2=\widehat{G}^{2\hat{r}}\frac{A'}{W}\frac{\pd W}{\pd \widehat{\Phi}^{\hat{r}}}+\frac{\kappa e^{-A}}{\ell W}\left(2\textrm{sech}\phi_4\tanh\phi_3\tanh\phi_4\frac{\pd W}{\pd \phi_3}-2\textrm{sech}\phi_4\frac{\pd W}{\pd \phi_4}\right),\qquad \\
& &\phi'_3=\widehat{G}^{3\hat{r}}\frac{A'}{W}\frac{\pd W}{\pd \widehat{\Phi}^{\hat{r}}}+\frac{\kappa e^{-A}}{\ell W}\left(2\textrm{sech}\phi_2\textrm{sech}\phi_3\textrm{sech}\phi_4\frac{\pd W}{\pd \phi_1}\right.\nonumber \\
& &\phantom{\phi'_3=}\left. -2\textrm{sech}\phi_4\tanh\phi_3\tanh\phi_4\frac{\pd W}{\pd \phi_2}+2\textrm{sech}\phi_4\tanh\phi_2\tanh\phi_3\frac{\pd W}{\pd \phi_4}\right),\\
& &\phi'_4=\widehat{G}^{4\hat{r}}\frac{A'}{W}\frac{\pd W}{\pd \widehat{\Phi}^{\hat{r}}}+\frac{\kappa e^{-A}}{\ell W}\left( 2\textrm{sech}\phi_4\frac{\pd W}{\pd \phi_2}-2\textrm{sech}\phi_4\tanh\phi_2\tanh\phi_3\frac{\pd W}{\pd \phi_3}\right)\qquad
\end{eqnarray}
with $\widehat{\Phi}^{\hat{r}}=(\phi_1,\phi_2,\phi_3,\phi_4)$. Before giving the solutions, we first note that in the limit $\ell \rightarrow \infty$, these equations reduce to the BPS equations for RG flows studied in \cite{N4_omega_flow} as expected. Furthermore, for $\phi_2=\phi_4=0$ or $\phi_1=\phi_3=0$, we recover the BPS equations for Janus solutions with $SO(2)\times SO(2)\times SO(2)\times SO(3)$ or $SO(2)\times SO(2)\times SO(3)\times SO(2)$ symmetries studied in \cite{N4_Janus}. 

\begin{figure}
  \centering
  \begin{subfigure}[b]{0.45\linewidth}
    \includegraphics[width=\linewidth]{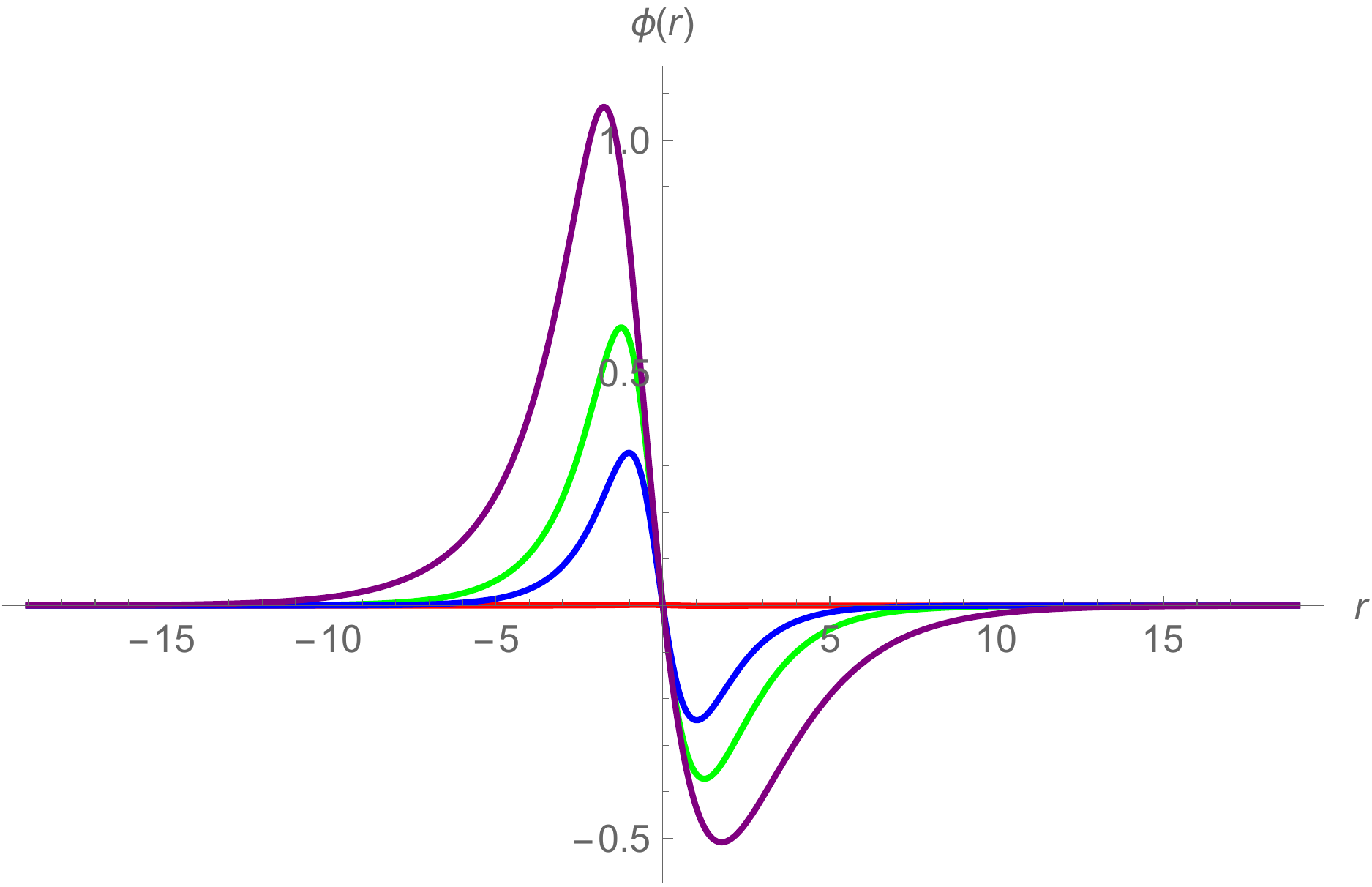}
  \caption{$\phi(r)$ solution}
  \end{subfigure}
  \begin{subfigure}[b]{0.45\linewidth}
    \includegraphics[width=\linewidth]{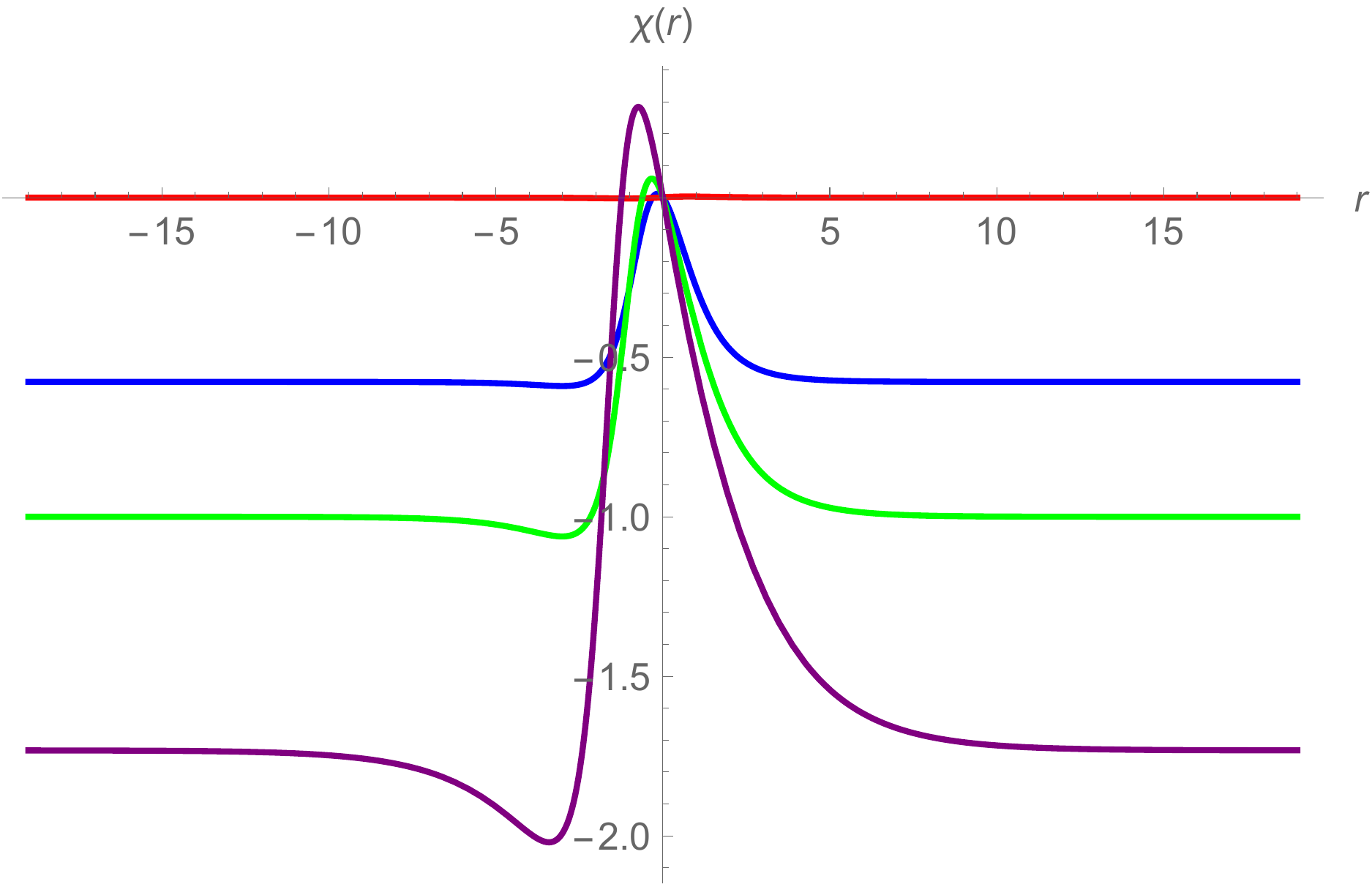}
  \caption{$\chi(r)$ solution}
  \end{subfigure}\\
    \begin{subfigure}[b]{0.45\linewidth}
    \includegraphics[width=\linewidth]{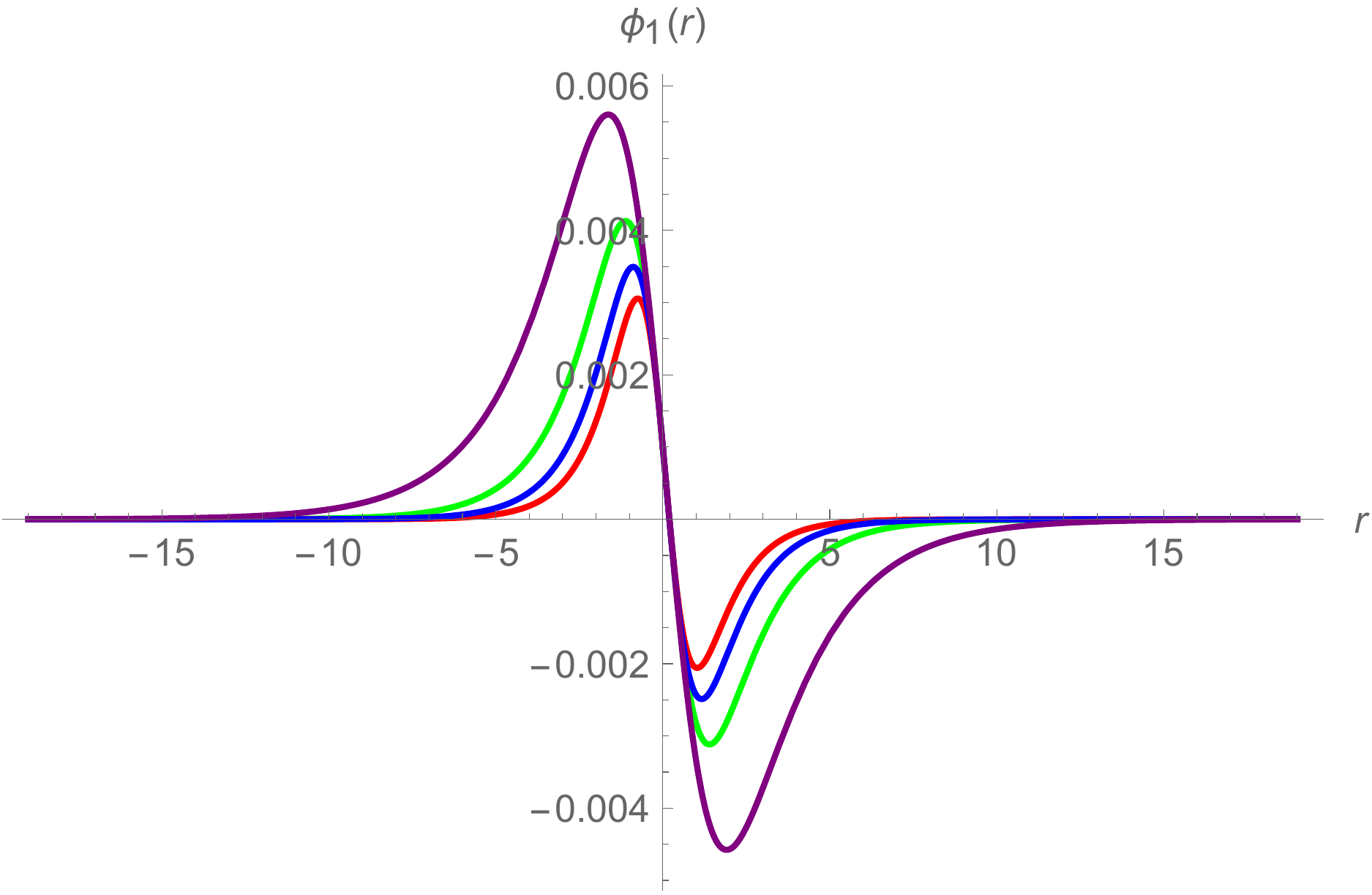}
  \caption{$\phi_1(r)$ solution}
  \end{subfigure}
  \begin{subfigure}[b]{0.45\linewidth}
    \includegraphics[width=\linewidth]{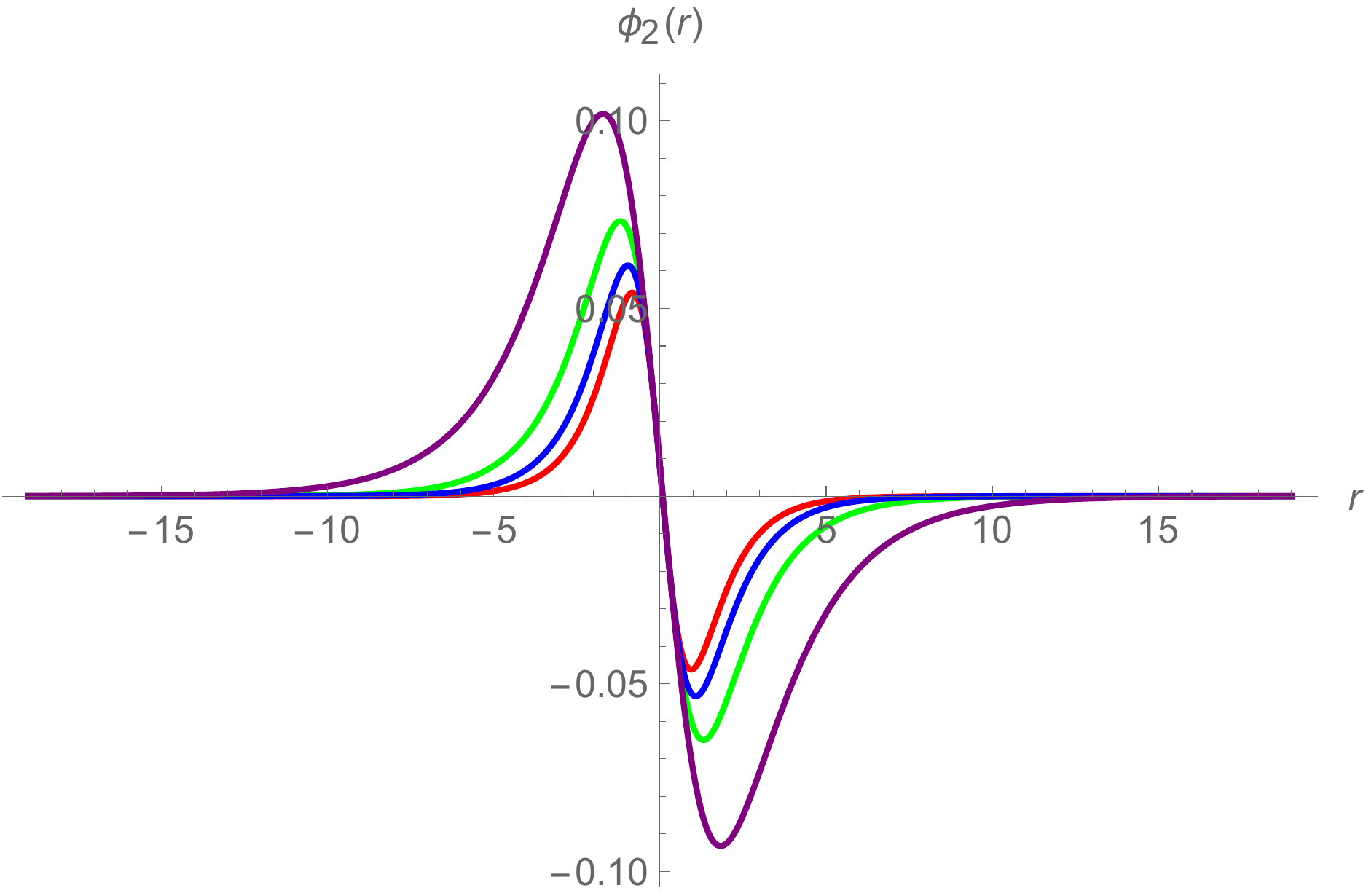}
  \caption{$\phi_2(r)$ solution}
  \end{subfigure}\\
  \begin{subfigure}[b]{0.45\linewidth}
    \includegraphics[width=\linewidth]{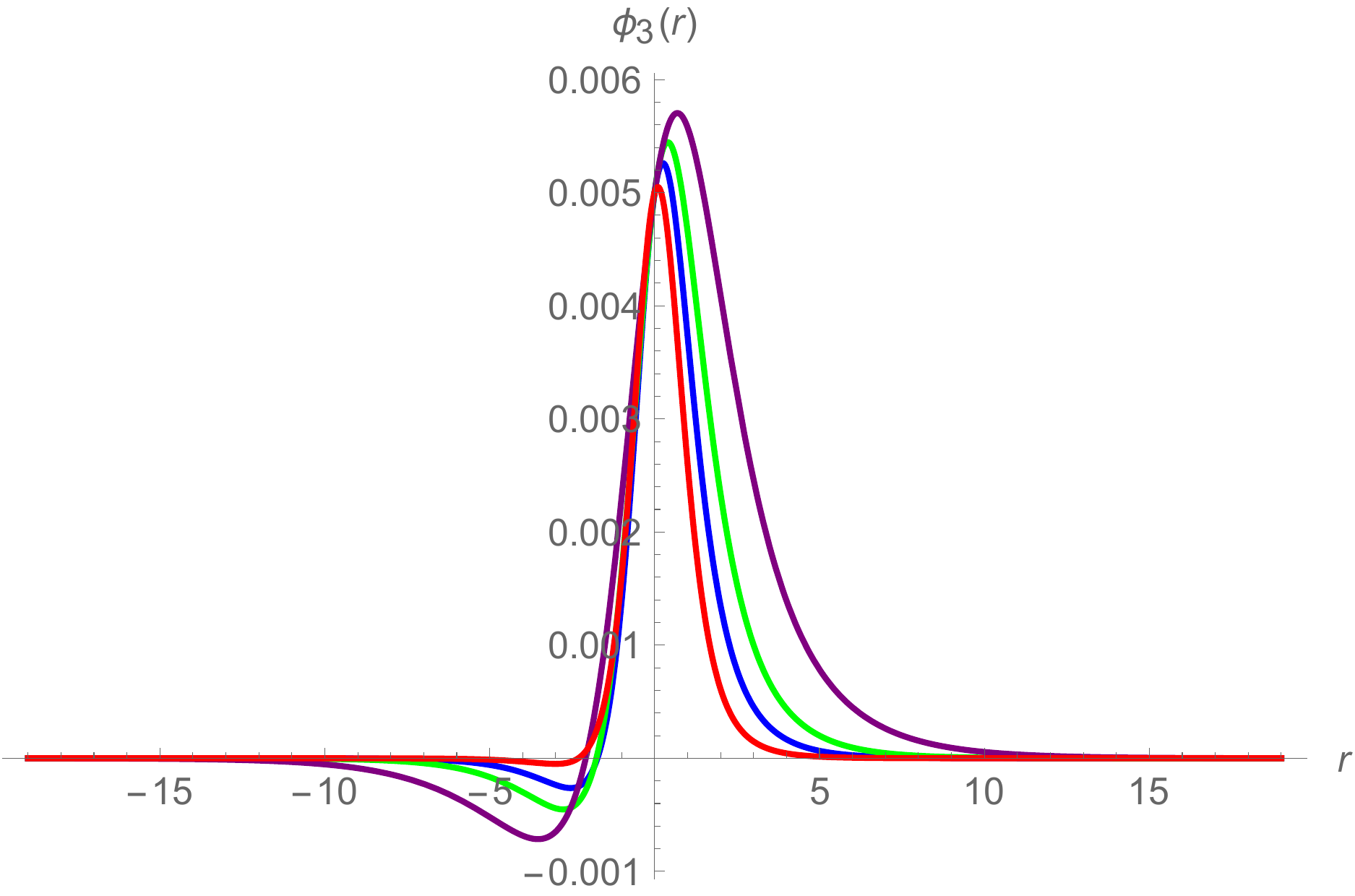}
  \caption{$\phi_3(r)$ solution}
  \end{subfigure}
  \begin{subfigure}[b]{0.45\linewidth}
    \includegraphics[width=\linewidth]{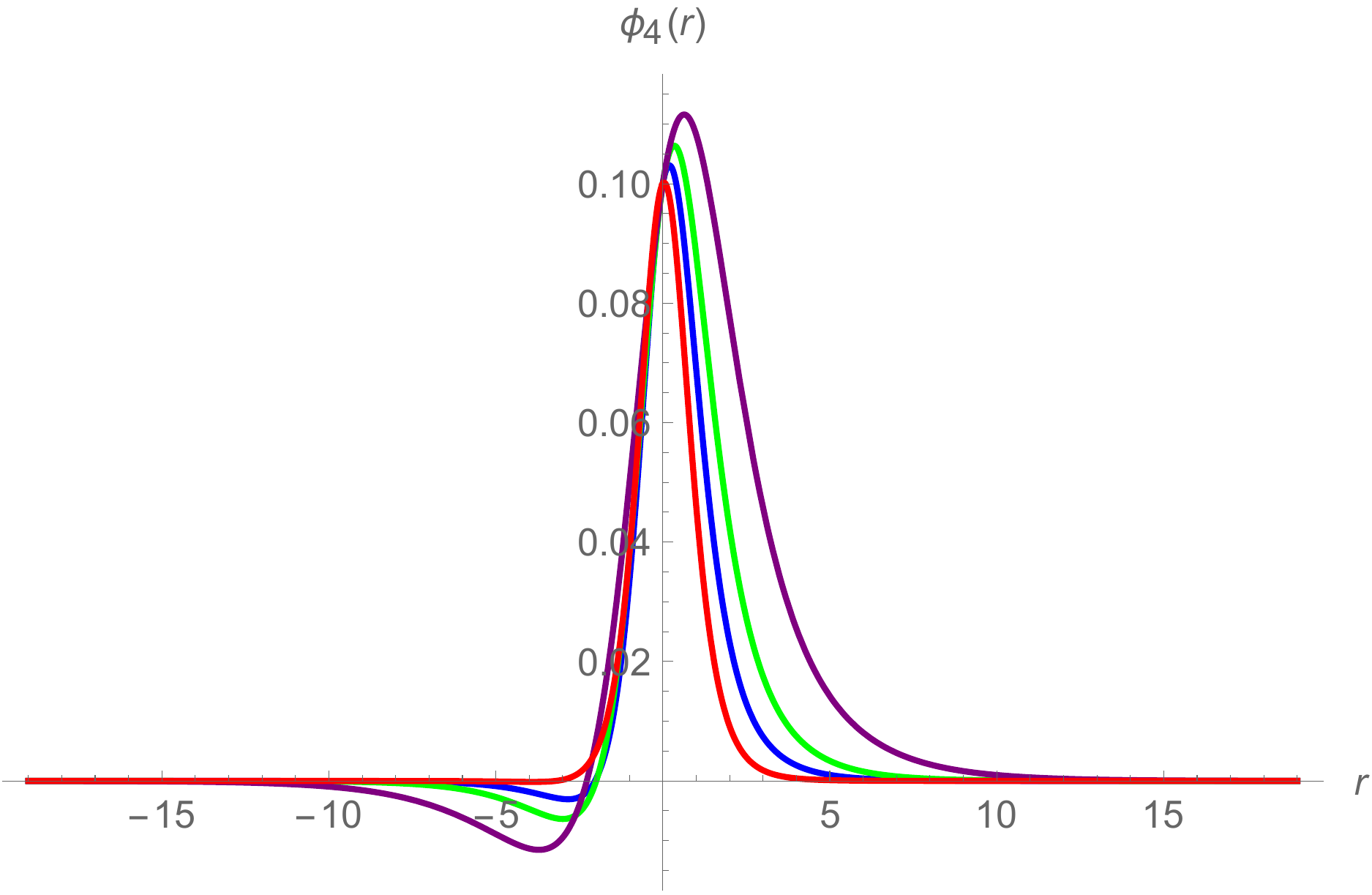}
  \caption{$\phi_4(r)$ solution}
  \end{subfigure}\\
   \begin{subfigure}[b]{0.45\linewidth}
    \includegraphics[width=\linewidth]{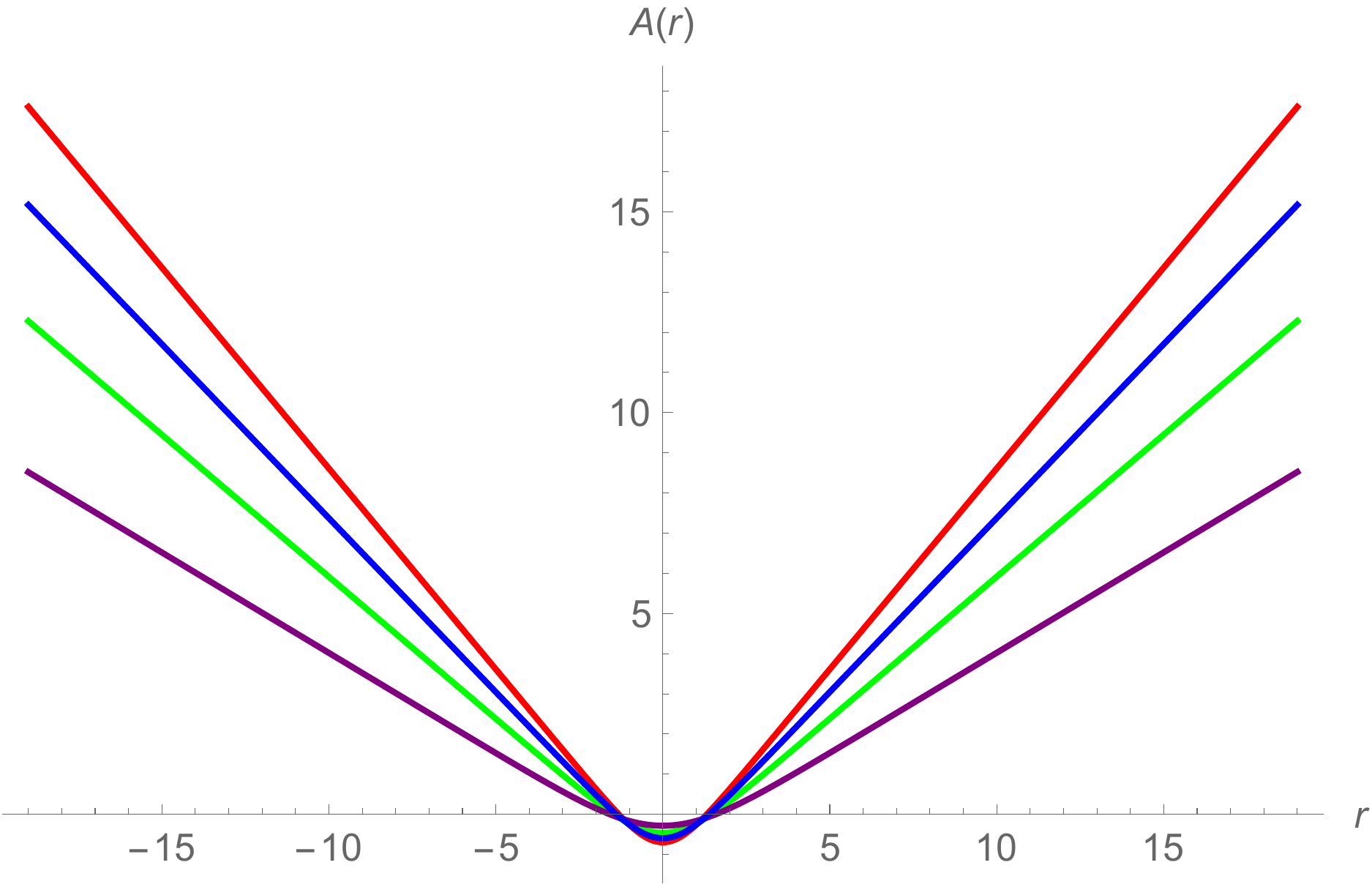}
  \caption{$A(r)$ solution}
   \end{subfigure}  
  \caption{Examples of $N=2$ Janus solutions interpolating between $N=4$ $AdS_4$ critical points with $SO(4)\times SO(4)$ symmetry for $g=1$, $\kappa=1$, $\ell=1$, $g_0=-g\sin\alpha$ and $\alpha=\frac{\pi}{6}$ (purple), $\alpha=\frac{\pi}{4}$ (green), $\alpha=\frac{\pi}{3}$ (blue), $\alpha=\frac{\pi}{2}$ (red).}
  \label{Fig1}
\end{figure}

We now give $N=2$ supersymmetric Janus solutions with $SO(2)\times SO(2)\times SO(2)\times SO(2)$ symmetry. After numerically solve the BPS equations, we find examples of Janus solutions for $g=1$, $\kappa=1$, $\ell=1$ and $g_0=-g\sin\alpha$ as in figure \ref{Fig1}. In the figure, we have depicted the solutions for different values of the phase $\alpha$. We also emphasize here that all values of $\alpha$ are equivalent to $\alpha=\frac{\pi}{2}$. We have given the solutions for various values of $\alpha$ only for clarity of the presentation since solutions with different boundary conditions but the same value of $\alpha$ are very close to each other and difficult to see. These solutions interpolate between $SO(4)\times SO(4)$ critical points and describe two-dimensional conformal defects within the $N=4$ SCFT. The defects are invariant under $SO(2)\times SO(2)\times SO(2)\times SO(2)$ subgroup of the $SO(4)\times SO(4)$ symmetry of the three-dimensional SCFT and preserve $N=(2,0)$ or $N=(0,2)$ supersymmetry in two dimensions depending on the values of $\kappa=1$ or $\kappa=-1$. 

\section{$N=1$ supersymmetric Janus solutions}\label{N1Janus}
We now move to $SO(3)_{\textrm{diag}}\times SO(3)$ sector which is a subtruncation of $SO(3)_{\textrm{diag}}$ sector studied in \cite{N4_omega_flow}. We will follow the notation of \cite{N4_omega_flow} for the sake of comparison. The $SO(3)_{\textrm{diag}}\times SO(3)$ sector contains two singlet scalars from $SO(6,6)/SO(6)\times SO(6)$ coset, see more detail in \cite{N4_omega_flow}, with the coset representative      
\begin{equation}
\mc{V}=e^{\phi_1\hat{Y}_1}e^{\phi_3\hat{Y}_3}
\end{equation}
in which the non-compact generators are given by
\begin{equation}
\hat{Y}_1=Y_{11}+Y_{22}+Y_{33}+Y_{44}\qquad \textrm{and}\qquad \hat{Y}_3=Y_{51}+Y_{62}+Y_{73}+Y_{84}\, .
\end{equation}
\indent The $A^{ij}_1$ tensor takes the form      
\begin{equation}
A^{ij}_1=\textrm{diag}(\mc{A},\mc{B},\mc{B},\mc{B})\label{A1_SO3diag}
\end{equation}
with $\mc{A}$ leading to the superpotential
\begin{eqnarray}
\mc{W}&=&\frac{1}{2}e^{\frac{\phi}{2}}\left[g\cosh^3\phi_3+h_1\sin\beta_1(i\sinh\phi_1-\cosh\phi_1\sinh\phi_3)^3\right]\nonumber \\
& &+\frac{1}{2}e^{-\frac{\phi}{2}}\left[g(\cosh\phi_1+i\sinh\phi_1\sinh\phi_3)^3-(\sinh\phi_1+i\cosh\phi_1\sinh\phi_3)^3\times\right.\nonumber \\
& &\left. h_1\cos\beta_1\right]+\frac{1}{2}e^{-\frac{\phi}{2}}\left[ig\cosh^3\phi_3+h_1\sin\beta_1(\sinh\phi_1+i\cosh\phi_1\sinh\phi_3)^3\right]\chi\, .\nonumber \\
& &\label{SO3d_SO3_W}
\end{eqnarray}
The solutions in this sector then preserve $N=1$ supersymmetry. To simplify the expressions, in this case, we will set $\alpha=\frac{\pi}{2}$ and $g_0=-g$.

For completeness, we also note that the scalar potential can be written as
\begin{equation}
V=4\left(\frac{\pd W}{\pd \phi}\right)^2+4e^{2\phi}\left(\frac{\pd W}{\pd \chi}\right)^2+\frac{2}{3}\textrm{sech}^2\phi_3\left(\frac{\pd W}{\pd \phi_1}\right)^2+\frac{2}{3}\left(\frac{\pd W}{\pd \phi_3}\right)^2-3W^2\, .
\end{equation}
The explicit form of this potential can be found in \cite{N4_omega_flow}. In this paper, we simply recall that the scalar potential admits three supersymmetric $AdS_4$ critical points. The first one is the trivial $SO(4)\times SO(4)$ critical point at which all scalars vanish for $\alpha=\frac{\pi}{2}$ and $g_0=-g$ while the other two are given by
 \begin{eqnarray}
i:\,\,\,\beta_1=0;\qquad & &\phi_3=\chi=0,\qquad \phi_1=\frac{1}{2}\ln\left[\frac{h_1+g}{h_1-g}\right],\nonumber \\
& &\phi=-\frac{1}{2}\ln\left[1-\frac{g^2}{h_1^2}\right],\qquad V_0=-\frac{3g^2h_1}{\sqrt{h_1^2-g^2}},\\
ii:\,\,\,\beta_1=\frac{\pi}{2};\qquad & &\phi_1=\chi=0,\qquad \phi_3=\frac{1}{2}\ln\left[\frac{h_1+g}{h_1-g}\right],\nonumber \\
& &\phi=\frac{1}{2}\ln\left[1-\frac{g^2}{h_1^2}\right],\qquad V_0=-\frac{3g^2h_1}{\sqrt{h_1^2-g^2}}\, .
\end{eqnarray}
Both of these critical points preserve $N=4$ supersymmetry as can be verified by setting $\chi=\phi_1=0$ or $\chi=\phi_3=0$ which gives $\mc{A}=\mc{B}$. On the other hand, for $\phi_1\neq 0$ and $\phi_3\neq 0$, the supersymmetry is broken to $N=1$. The holographic RG flows between these critical points preserving $N=4$ and $N=1$ supersymmetries have already been studied in \cite{N4_omega_flow}. 

In the present work, we are interested in supersymmetric Janus solutions. First of all, we note that setting either $\phi_1=0$ or $\phi_3=0$ does not lead to a consistent set of BPS equations for Janus solutions. This implies that, unlike the RG flow case, there are no $N=4$ supersymmetric Janus solutions with $SO(3)_{\textrm{diag}}\times SO(3)\times SO(3)$ or $SO(3)\times SO(3)_{\textrm{diag}}\times SO(3)$ symmetries. For $\phi_1\neq 0$ and $\phi_3\neq 0$, truncating out $\chi$ is also not consistent with the BPS equations. Therefore, $N=1$ Janus solutions must involve all scalars in the $SO(3)_{\textrm{diag}}\times SO(3)$ sector as in the case of $N=1$ RG flow solutions found in \cite{N4_omega_flow}.
 
\begin{figure}
  \centering
  \begin{subfigure}[b]{0.45\linewidth}
    \includegraphics[width=\linewidth]{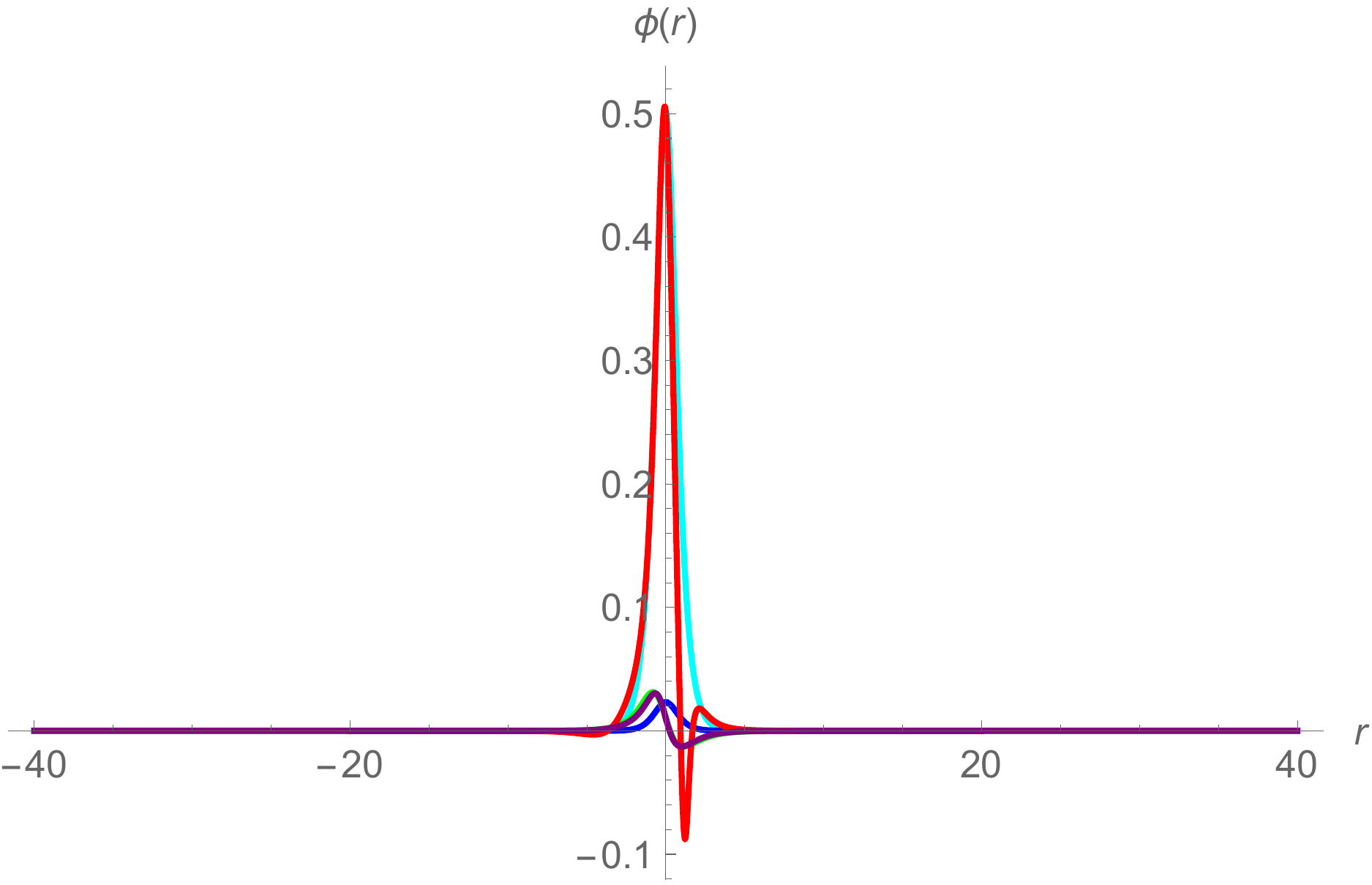}
  \caption{$\phi(r)$ solution}
  \end{subfigure}
  \begin{subfigure}[b]{0.45\linewidth}
    \includegraphics[width=\linewidth]{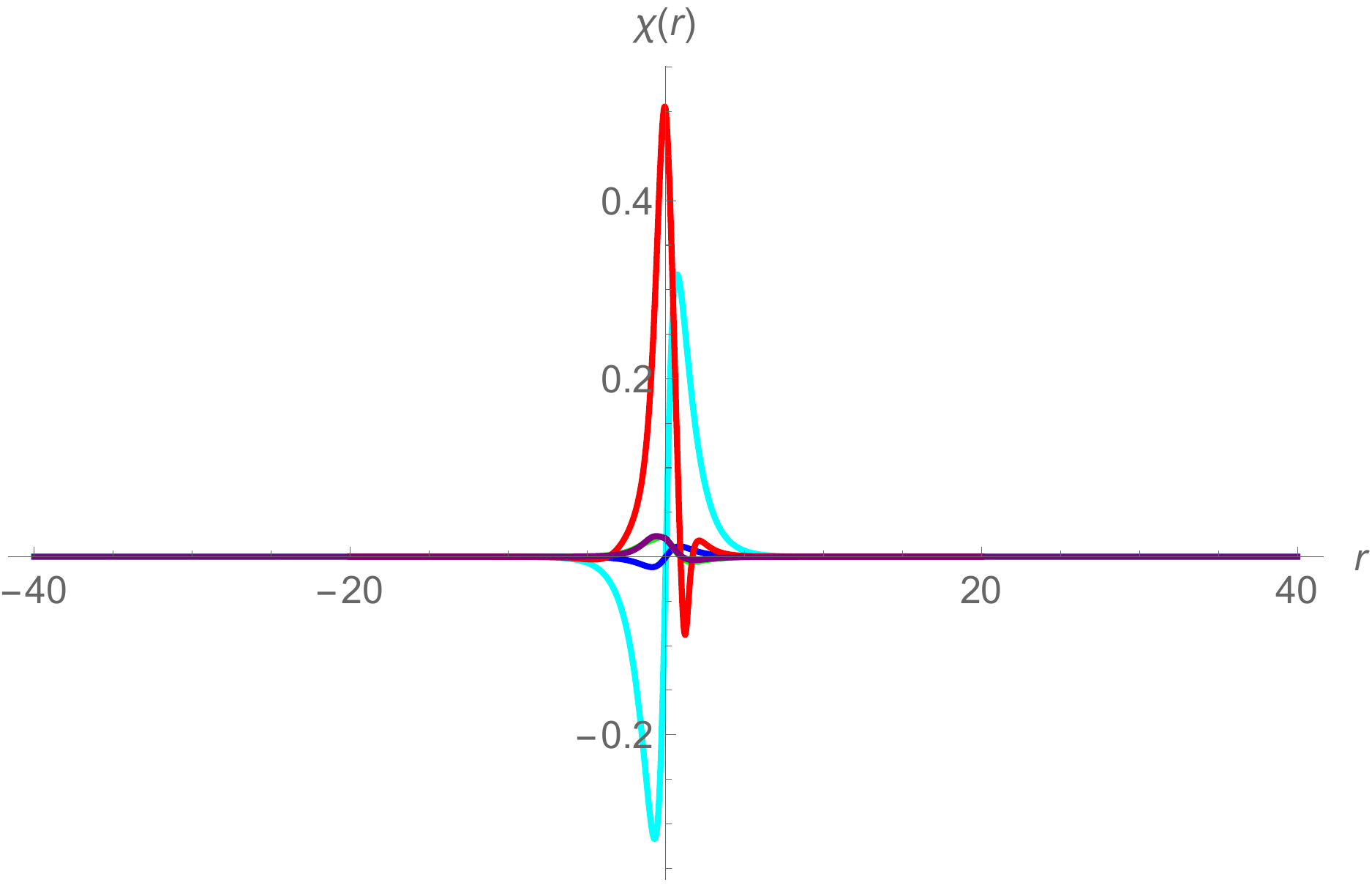}
  \caption{$\chi(r)$ solution}
  \end{subfigure}\\
    \begin{subfigure}[b]{0.45\linewidth}
    \includegraphics[width=\linewidth]{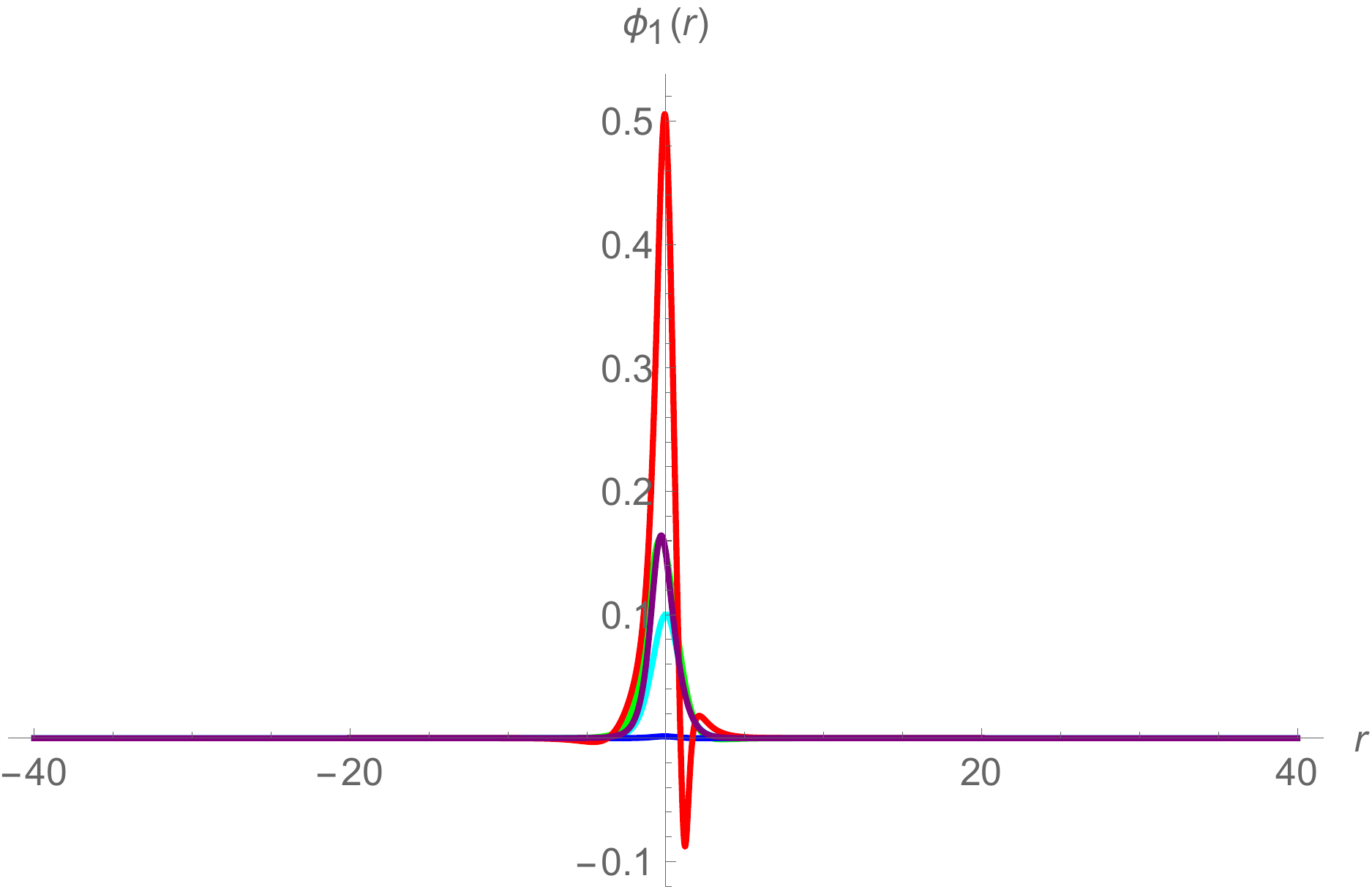}
  \caption{$\phi_1(r)$ solution}
  \end{subfigure}
  \begin{subfigure}[b]{0.45\linewidth}
    \includegraphics[width=\linewidth]{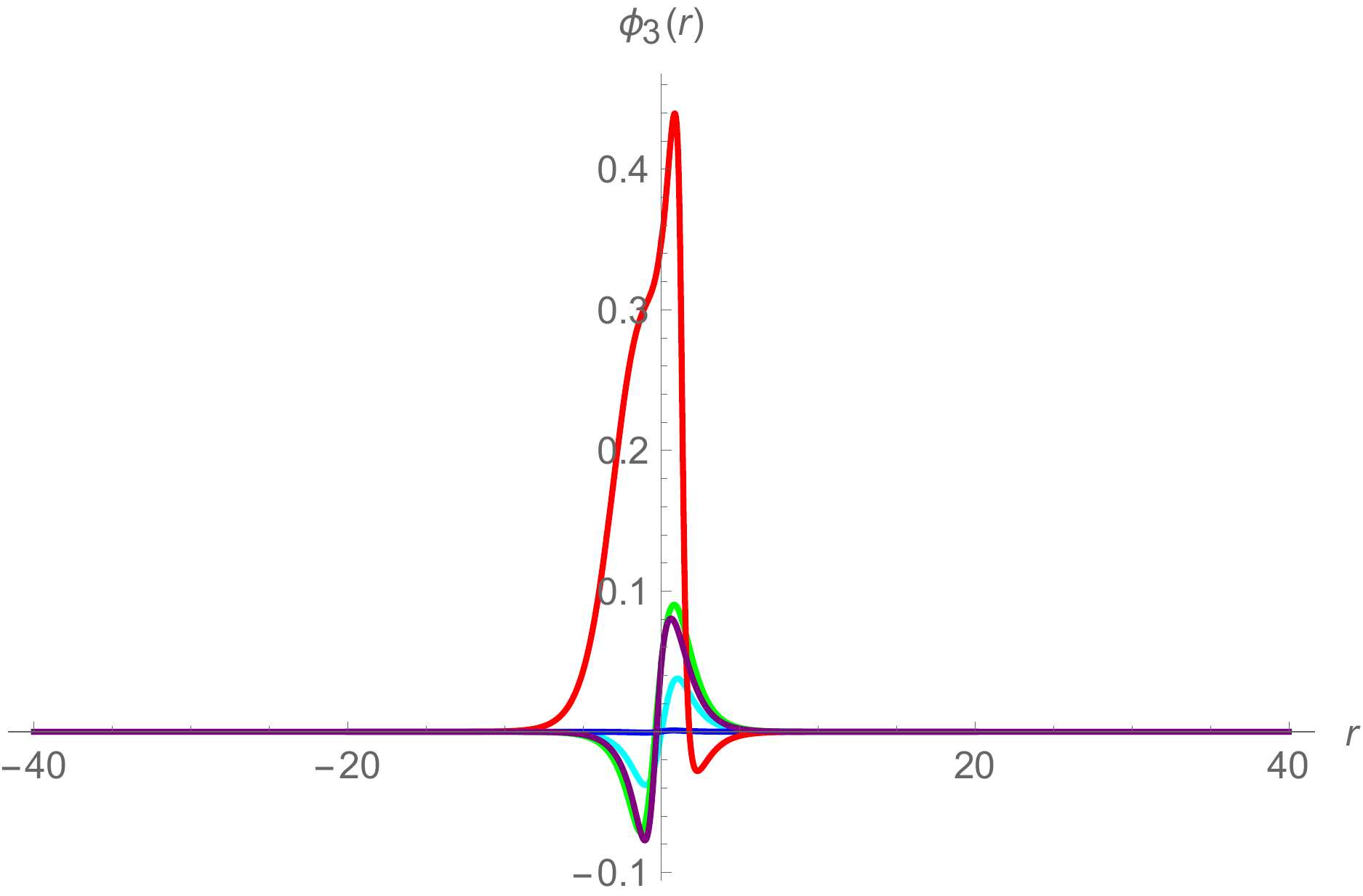}
  \caption{$\phi_3(r)$ solution}
  \end{subfigure}\\
   \begin{subfigure}[b]{0.45\linewidth}
    \includegraphics[width=\linewidth]{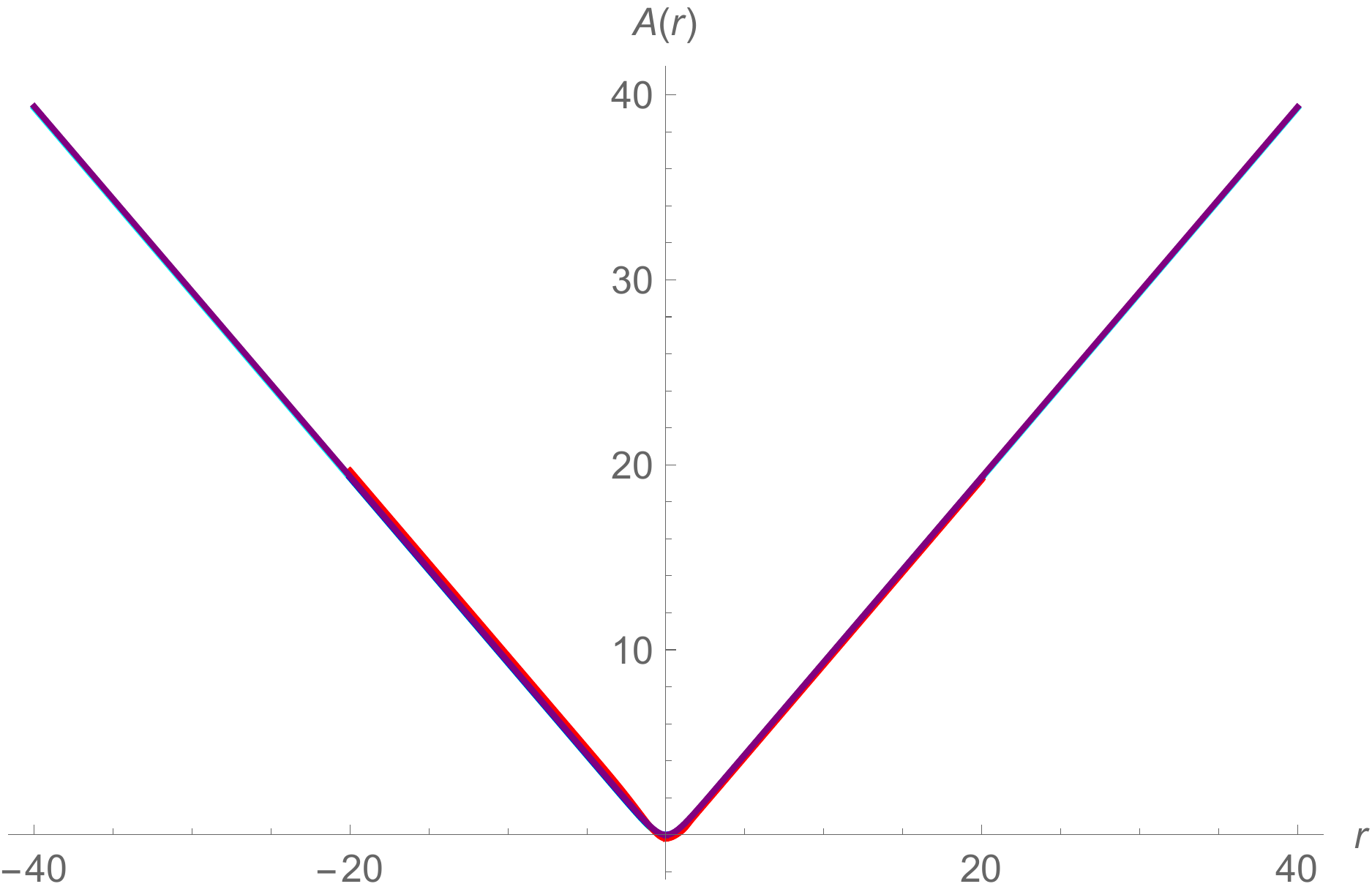}
  \caption{$A(r)$ solution}
   \end{subfigure}  
  \caption{Examples of $N=1$ Janus solutions interpolating between $N=4$ $AdS_4$ critical points with $SO(4)\times SO(4)$ symmetry for different values of $\beta_1$, $\beta_1=0$ (cyan), $\beta_1=\frac{\pi}{6}$ (red), $\beta_1=\frac{\pi}{4}$ (green), $\beta_1=\frac{\pi}{3}$ (blue), $\beta_1=\frac{\pi}{2}$ (purple).}
  \label{Fig2}
\end{figure}

By the same procedure as in the previous section with $\epsilon^{2}=\epsilon^{3}=\epsilon^{4}=0$, we find that the BPS equations can be written as
 \begin{eqnarray}
 \phi'&=&-4\frac{A'}{W}\frac{\pd W}{\pd \phi}-4e^\phi\frac{e^{-A}\kappa}{\ell W}\frac{\pd W}{\pd \chi},\\
 \chi'&=&-4e^{2\phi}\frac{A'}{W}\frac{\pd W}{\pd \phi}+4e^\phi\frac{e^{-A}\kappa}{\ell W}\frac{\pd W}{\pd \phi},\\
 \phi'_1&=&-\frac{2}{3}\textrm{sech}^2\phi_3\frac{A'}{W}\frac{\pd W}{\pd \phi_1}-\frac{2}{3}\textrm{sech}\phi_3\frac{e^{-A}\kappa}{\ell W}\frac{\pd W}{\pd \phi_3},\\
 \phi'_3&=&-\frac{2}{3}\phi_3\frac{A'}{W}\frac{\pd W}{\pd \phi_3}+\frac{2}{3}\textrm{sech}\phi_3\frac{e^{-A}\kappa}{\ell W}\frac{\pd W}{\pd \phi_1}
 \end{eqnarray}
together with the usual equation for the metric function
\begin{equation}
{A'}^2+\frac{e^{-2A}}{\ell^2}=W^2
\end{equation}
with the superpotential given in \eqref{SO3d_SO3_W}. It should be noted that these equations again reduce to the BPS equations for holographic RG flows studied in \cite{N4_omega_flow} in the limit $\ell\rightarrow \infty$ as expected.
\\
\indent We begin with generic solutions for different values of the phase $\beta_1$. After numerically solve the BPS equations, we find examples of solutions for $g=1$, $h_1=2$, $\kappa=1$ and $\ell=1$ as shown in figure \ref{Fig2}. Some of the solutions are very close to each other, so some solution, in particular the one represented by the green line, is not clearly seen. From the figure, all the solutions are qualitatively similar and describe two-dimensional conformal defects within a three-dimensional $N=4$ SCFT with $SO(4)\times SO(4)$ symmetry. Unlike the solutions in the previous section, these defects preserve only $N=(1,0)$ or $N=(0,1)$ supersymmetry depending on the values of $\kappa=1$ or $\kappa=-1$.  

\begin{figure}
  \centering
  \begin{subfigure}[b]{0.45\linewidth}
    \includegraphics[width=\linewidth]{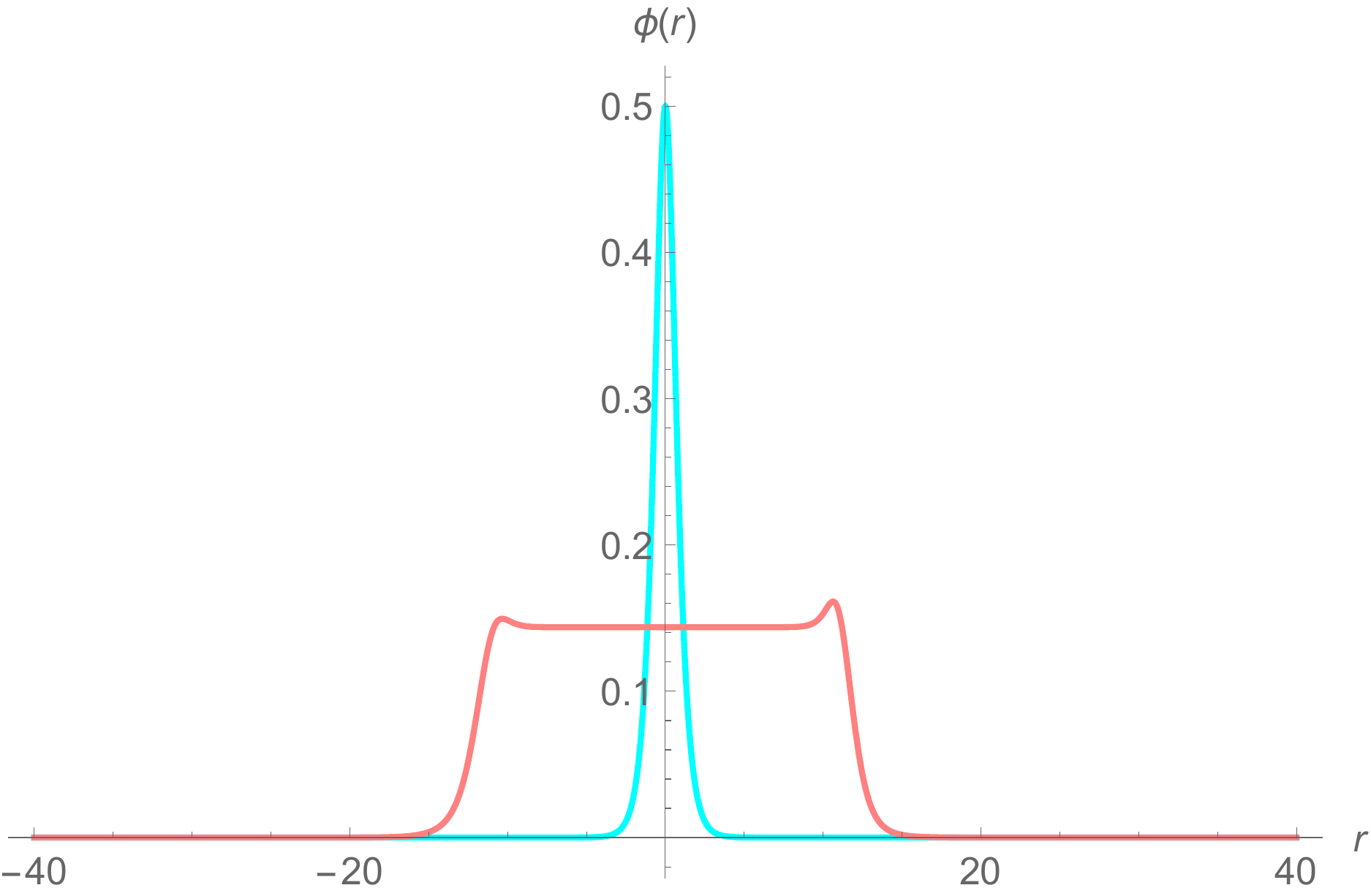}
  \caption{$\phi(r)$ solution}
  \end{subfigure}
  \begin{subfigure}[b]{0.45\linewidth}
    \includegraphics[width=\linewidth]{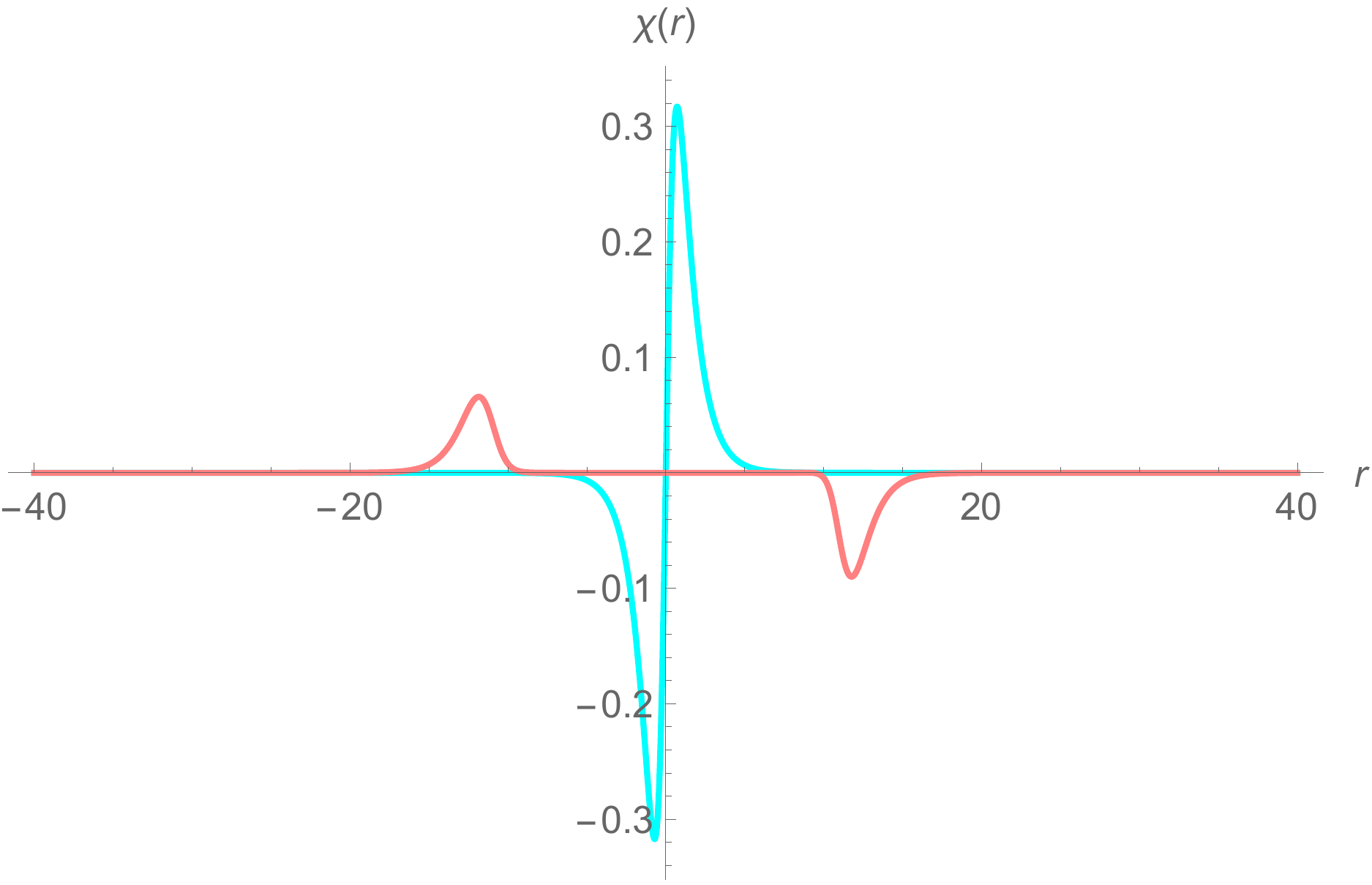}
  \caption{$\chi(r)$ solution}
  \end{subfigure}\\
    \begin{subfigure}[b]{0.45\linewidth}
    \includegraphics[width=\linewidth]{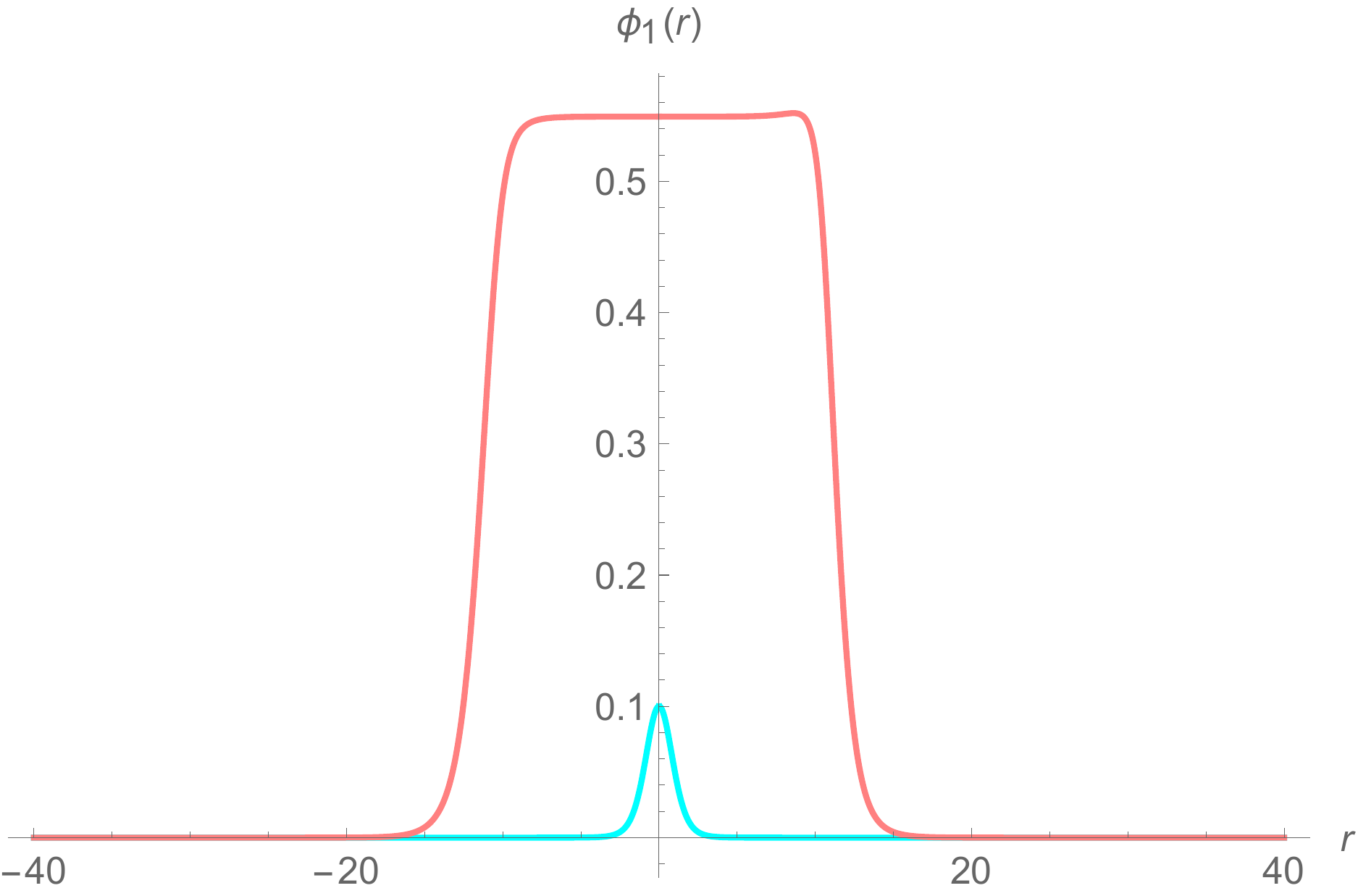}
  \caption{$\phi_1(r)$ solution}
  \end{subfigure}
  \begin{subfigure}[b]{0.45\linewidth}
    \includegraphics[width=\linewidth]{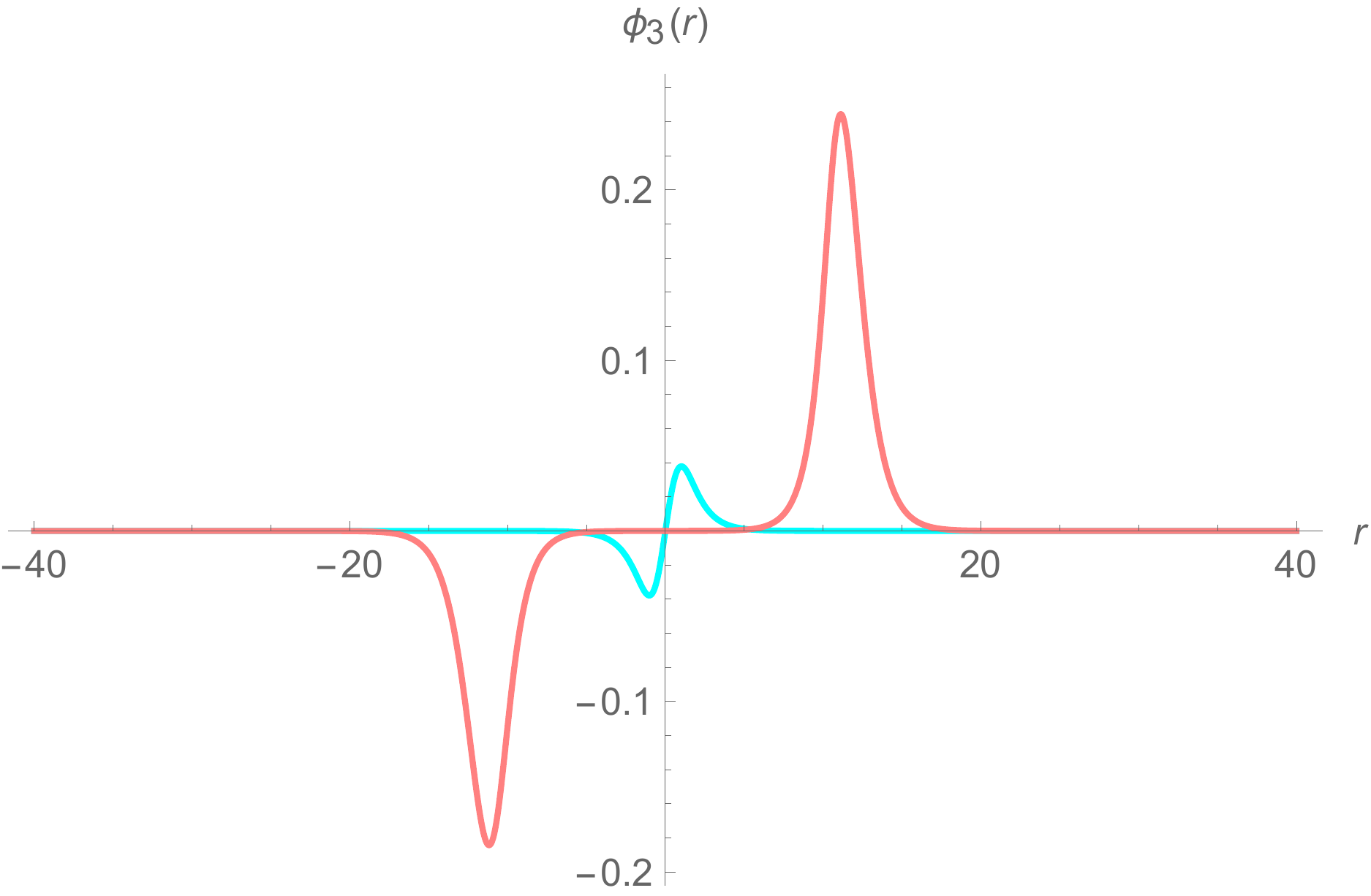}
  \caption{$\phi_3(r)$ solution}
  \end{subfigure}\\
   \begin{subfigure}[b]{0.45\linewidth}
    \includegraphics[width=\linewidth]{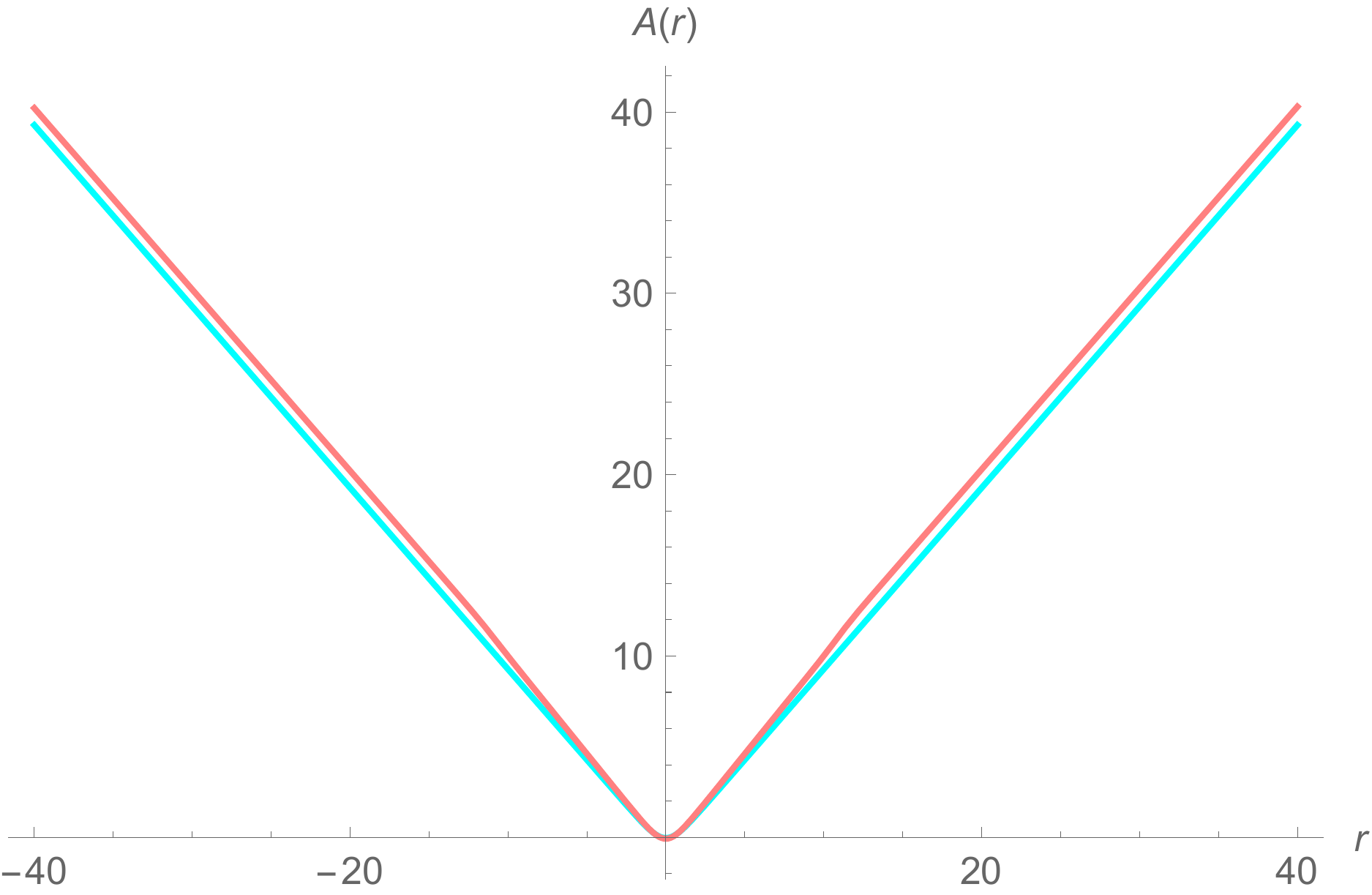}
  \caption{$A(r)$ solution}
   \end{subfigure}  
  \begin{subfigure}[b]{0.45\linewidth}
    \includegraphics[width=\linewidth]{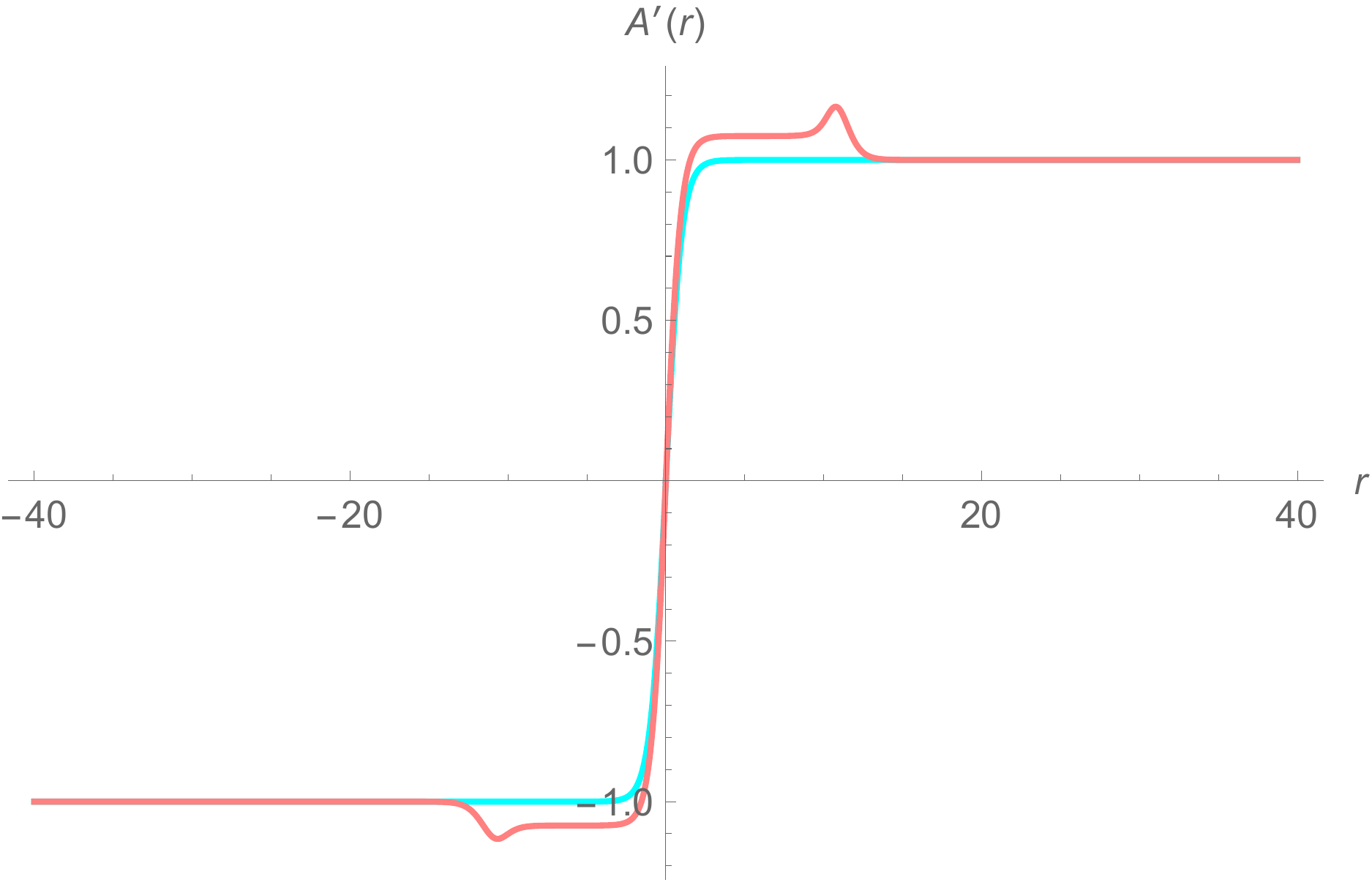}
  \caption{$A'(r)$ solution}
  \end{subfigure}
    \caption{An example of $N=1$ Janus solutions (pink) interpolating between $N=4$ $AdS_4$ critical points with $SO(3)_{\textrm{diag}}\times SO(3)\times SO(3)$ symmetry (critical point $i$).}
  \label{Fig3}
\end{figure}

\begin{figure}
  \centering
  \begin{subfigure}[b]{0.45\linewidth}
    \includegraphics[width=\linewidth]{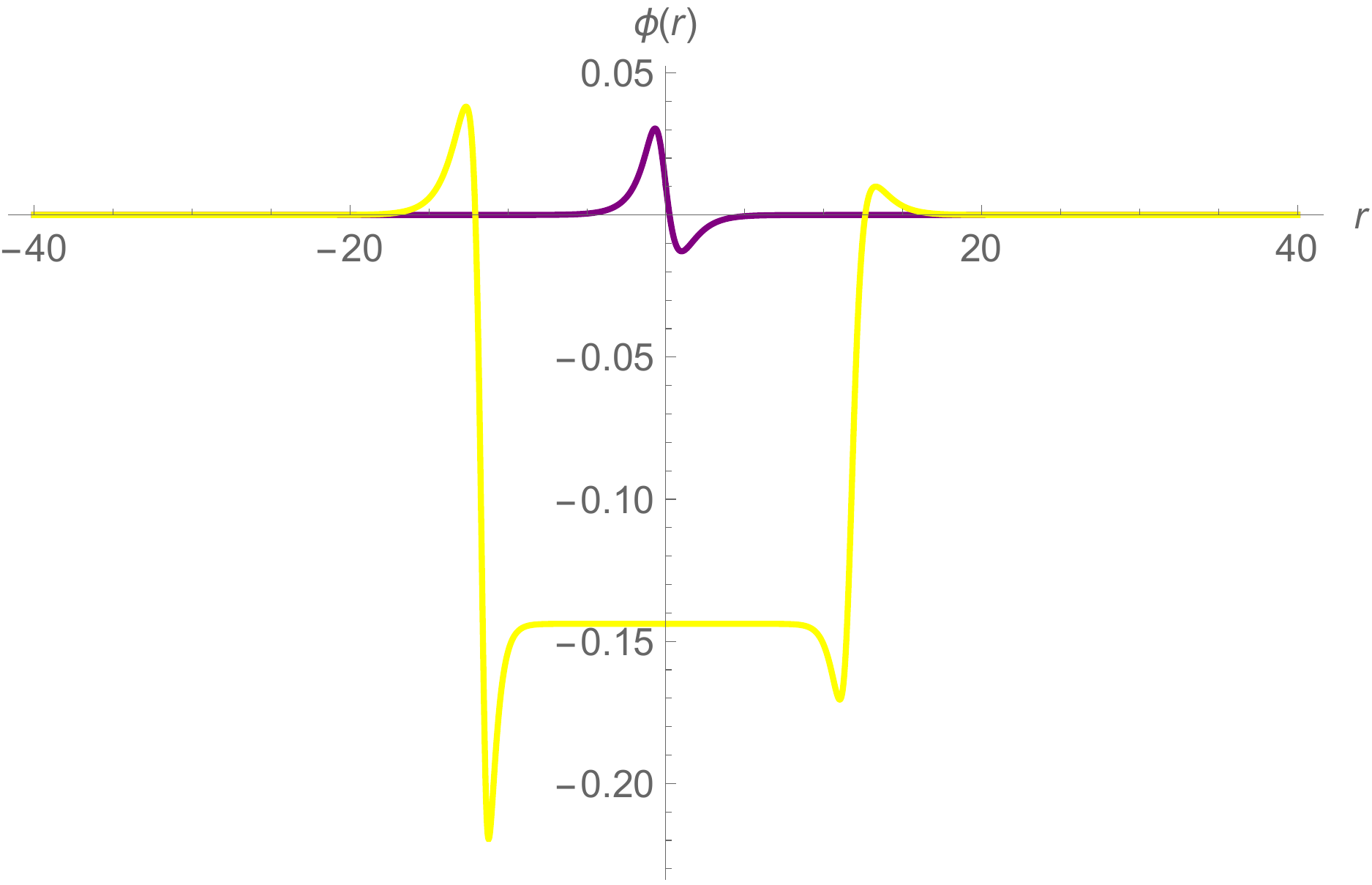}
  \caption{$\phi(r)$ solution}
  \end{subfigure}
  \begin{subfigure}[b]{0.45\linewidth}
    \includegraphics[width=\linewidth]{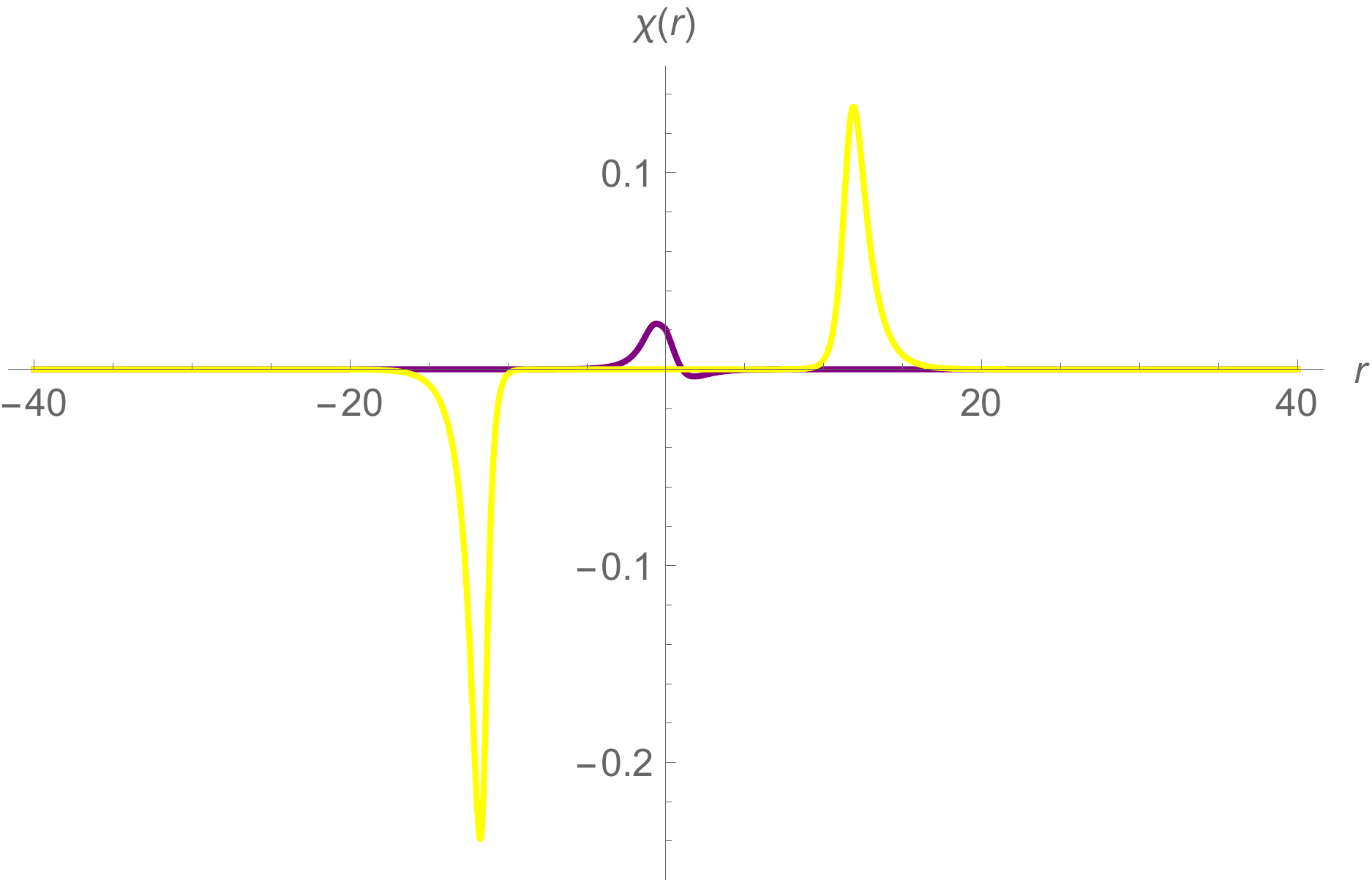}
  \caption{$\chi(r)$ solution}
  \end{subfigure}\\
    \begin{subfigure}[b]{0.45\linewidth}
    \includegraphics[width=\linewidth]{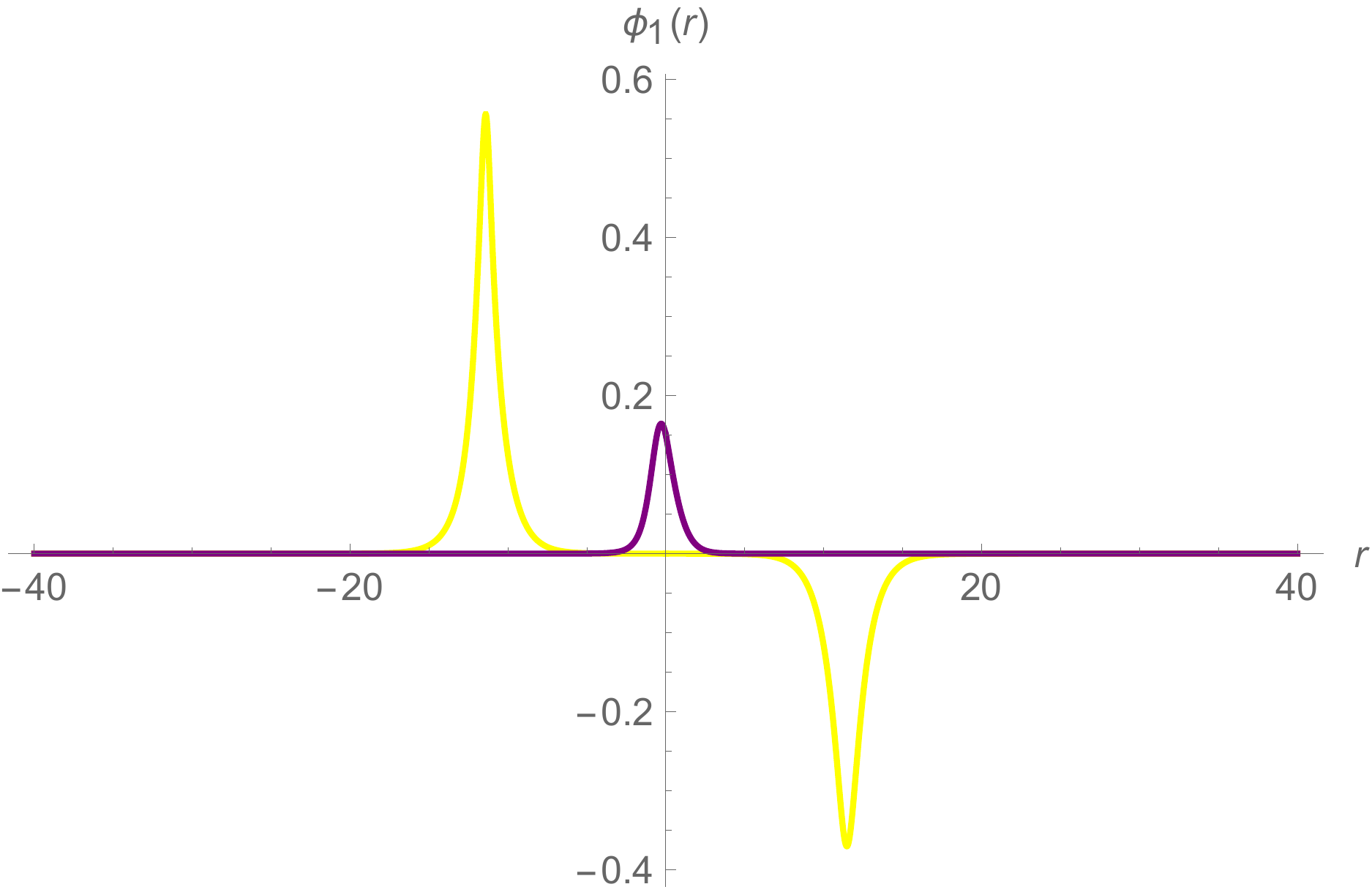}
  \caption{$\phi_1(r)$ solution}
  \end{subfigure}
  \begin{subfigure}[b]{0.45\linewidth}
    \includegraphics[width=\linewidth]{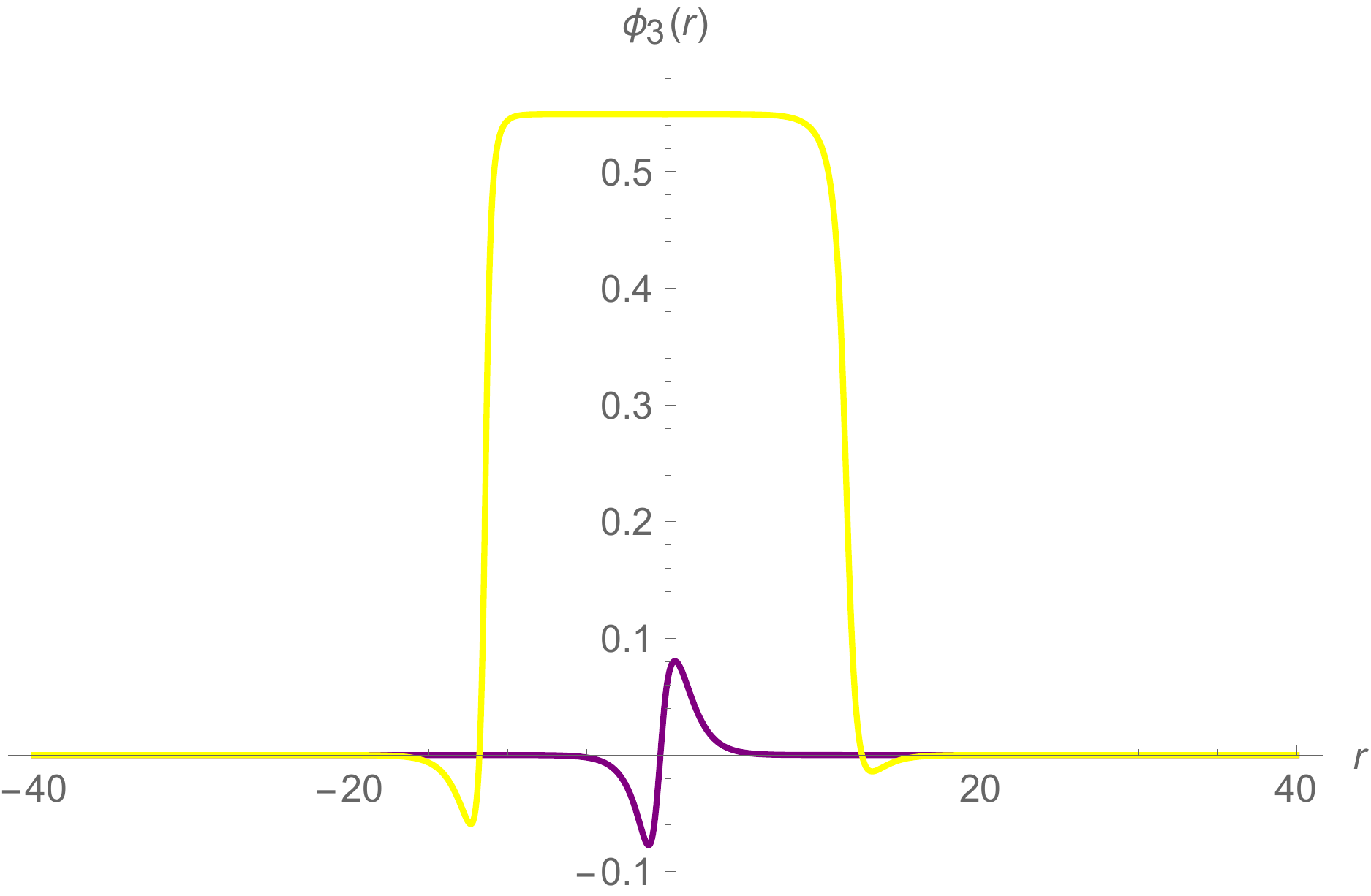}
  \caption{$\phi_3(r)$ solution}
  \end{subfigure}\\
   \begin{subfigure}[b]{0.45\linewidth}
    \includegraphics[width=\linewidth]{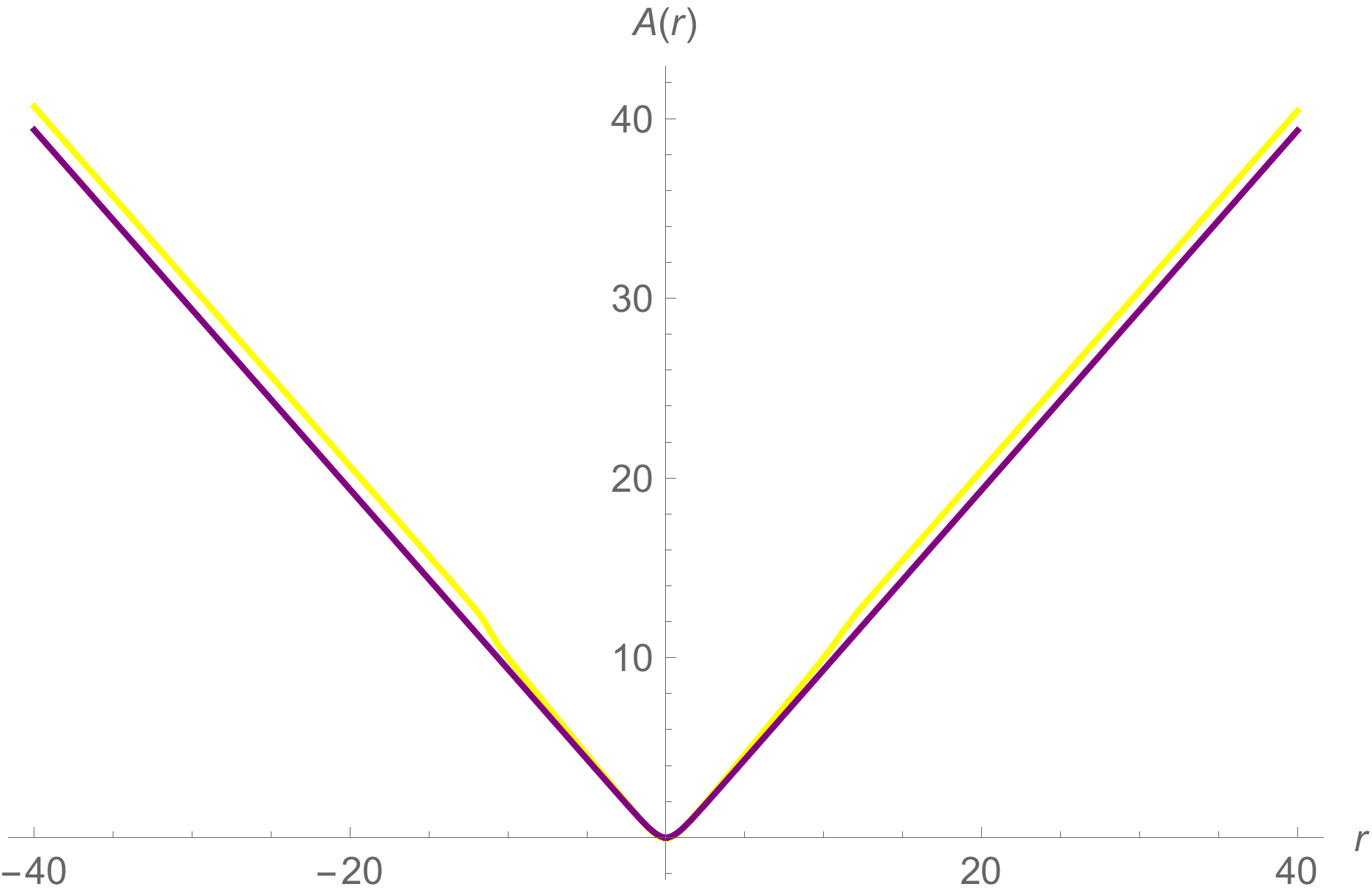}
  \caption{$A(r)$ solution}
   \end{subfigure}  
  \begin{subfigure}[b]{0.45\linewidth}
    \includegraphics[width=\linewidth]{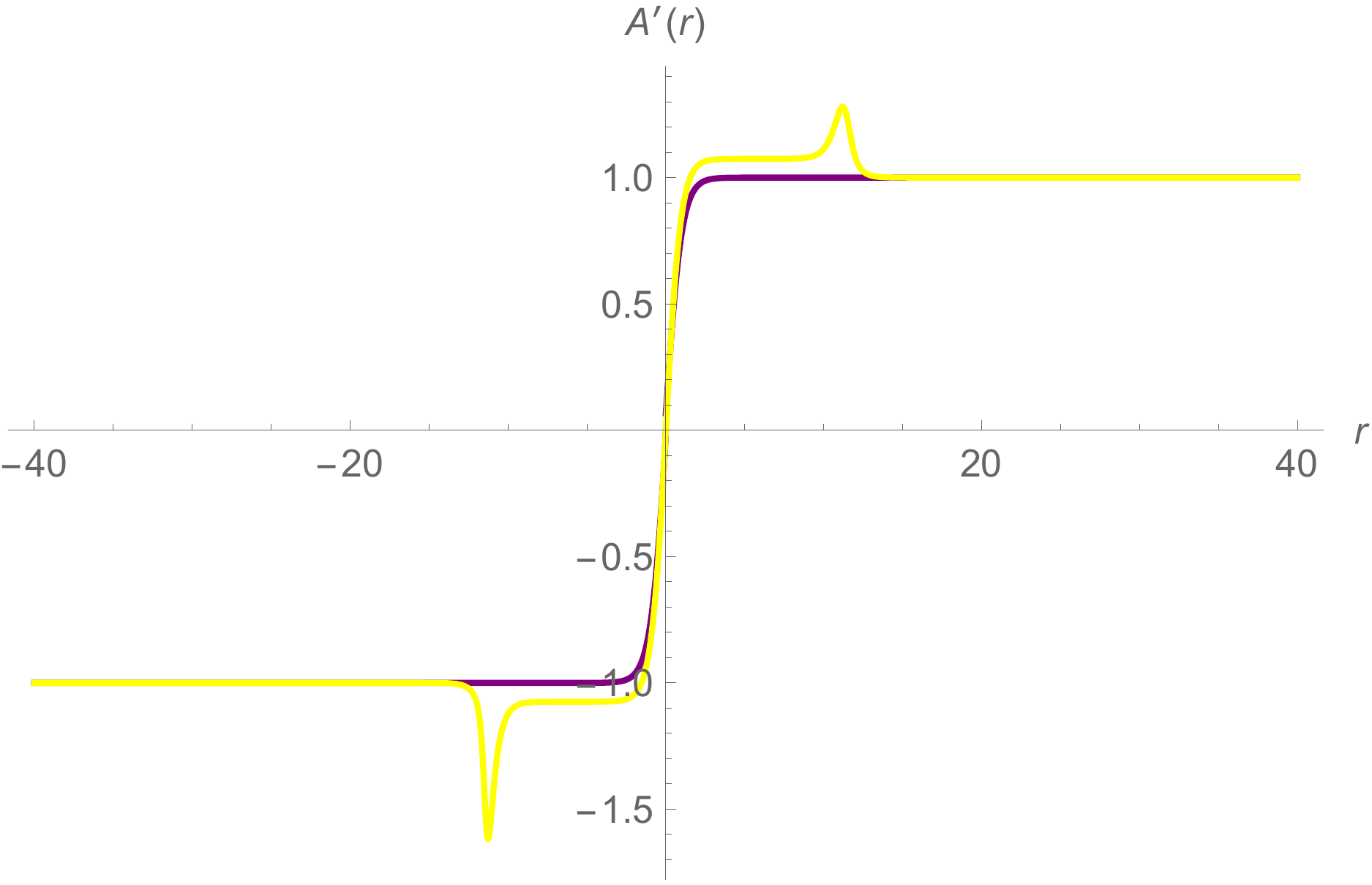}
  \caption{$A'(r)$ solution}
  \end{subfigure}
    \caption{An example of $N=1$ Janus solutions (yellow) interpolating between $N=4$ $AdS_4$ critical points with $SO(3)\times SO(3)_{\textrm{diag}}\times SO(3)$ symmetry (critical point $ii$).}
  \label{Fig4}
\end{figure}

For $\beta_1=0$, there are two $AdS_4$ critical points, the trivial one and critical point $i$. With appropriate boundary conditions, we find a Janus solution interpolating between critical point $i$ for $g=1$, $h_1=2$, $\kappa=1$ and $\ell=1$ as shown in figure \ref{Fig3}. This solution is represented by the pink line. We have included the Janus solution between $SO(4)\times SO(4)$ critical points (cyan) for comparison. We have also given the solution for $A'(r)$ to explicitly show that the two solutions indeed interpolate between different pairs of critical points. However, it should be noted that critical point $i$ on both sides is generated by a holographic RG flow from the $SO(4)\times SO(4)$ critical point. In particular, this RG flow is one of the solutions studied recently in \cite{N4_omega_flow}. The Janus solution is accordingly similar to those given in \cite{warner_Janus,Minwoo_4DN8_Janus,N8_omega_Janus,ISO7_Janus,3D_Janus2}. A similar solution interpolating between critical points $ii$ can also be found as shown by the yellow line in figure \ref{Fig4} with $g=1$, $h_1=2$, $\kappa=1$ and $\ell=1$. As in figure \ref{Fig3}, we have included the Janus solution between $SO(4)\times SO(4)$ critical points for comparison (purple line). These two solutions describe $N=(1,0)$ or $N=(0,1)$ conformal defects within $N=4$ SCFTs with $SO(3)_{\textrm{diag}}\times SO(3)\times SO(3)$ or $SO(3)\times SO(3)_{\textrm{diag}}\times SO(3)$ symmetry. 

\section{Conclusions and discussions}\label{conclusion}
In this paper, we have studied supersymmetric Janus solutions from four-dimensional $N=4$ gauged supergravity with $SO(4)\times SO(4)$ gauge group in the presence of symplectic deformations. We have found two classes of solutions preserving $N=1$ and $N=2$ supersymmetries. The $N=2$ solutions interpolate between the trivial $N=4$ critical points with $SO(4)\times SO(4)$ symmetry. In this case, electric-magnetic phases or deformation parameters do not appear apart from those fixed by $SL(2,\mathbb{R})$ transformations and redefinitions of the dilaton and axion, and there are no other $AdS_4$ critical points. The solutions are invariant under $SO(2)\times SO(2)\times SO(2)\times SO(2)$ symmetry and describe $N=(2,0)$ or $N=(0,2)$ two-dimensional conformal defects in the $N=4$ SCFT dual to the $AdS_4$ critical point. 
\\
\indent On the other hand, in the $N=1$ case, we have found more interesting solutions. The solutions are obtained in $SO(3)_{\textrm{diag}}\times SO(3)$ sector, and for particular values of the phase $\beta_1=0$ and $\beta_1=\frac{\pi}{2}$, there are two additional non-trivial $N=4$ critical points with $SO(3)_{\textrm{diag}}\times SO(3)\times SO(3)$ and $SO(3)\times SO(3)_{\textrm{diag}}\times SO(3)$ symmetries apart from the trivial critical point. There are $N=1$ solutions interpolating between $SO(4)\times SO(4)$ critical points for any values of the electric-magnetic phase $\beta_1$ as in the $N=2$ solutions. Moreover, we have found solutions interpolating between $SO(3)_{\textrm{diag}}\times SO(3)\times SO(3)$ critical points and between $SO(3)\times SO(3)_{\textrm{diag}}\times SO(3)$ critical points. In this case, the solutions describe two-dimensional conformal defects in $N=4$ SCFTs dual to $AdS_4$ critical points $i$ and $ii$ that preserve $N=(1,0)$ or $N=(0,1)$ supersymmetries on the defects. These are the first examples of Janus solutions in $N=4$ gauged supergavity that involve non-trivial $AdS_4$ critical points. 
\\
\indent It would be interesting to identify the $N=4$ SCFTs dual to the $AdS_4$ critical points considered here and study the conformal defects dual to the Janus solutions found in this paper. As pointed out in \cite{N4_omega_flow}, in the $SO(3)_{\textrm{diag}}$ invariant scalar sector, both of the electric-magnetic phases $\beta_1$ and $\beta_2$ appear in the scalar potential and the superpotential. It would be of particular interest to investigate this sector and look for new supersymmetric $AdS_4$ vacua and also find new Janus solutions in this case. Finally, since $SO(4)\times SO(4)$ gauged supergravity admitting $AdS_4$ vacua for any values of the deformation parameters presently has no known embedding in higher dimensions, it could be highly desirable to find the corresponding embedding that would provide an uplift for the solutions found here and those given in \cite{N4_Janus,N4_omega_flow,4D_N4_flows} to ten/eleven dimensions. Along this line, recent developments in double field theory formalism would be very useful, see for example \cite{Dibitetto_SL2_angle,Dibitetto_non_Geometric,7D_N2_Malek,EFT_Heterotic_Malek,Malek_half_SUSY_EFT,Henning_Malek_AdS7_6_pure,
Henning_Malek_AdS7_6}. The uplifted solutions should provide a complete gravity dual of the $N=4$ SCFTs in three dimensions together with deformations and conformal defects in string/M-theory context. We leave these issues for future work.           
\vspace{0.5cm}\\
{\large{\textbf{Acknowledgement}}} \\
This work is funded by National Research Council of Thailand (NRCT) and Chulalongkorn University under grant N42A650263.

\end{document}